\documentclass[useAMS,usenatbib]{mn2e}
\usepackage{natbib}
\usepackage{graphicx}
\usepackage{subfigure}
\usepackage{multirow}
\usepackage{amsmath}
\usepackage{amssymb}
\usepackage{epstopdf}
\usepackage{rotating}
\usepackage{soul}

\usepackage{txfonts}
\usepackage{times}

\title[SMBH merger host morphologies]{
Morphological evolution of supermassive black hole merger hosts and multimessenger signatures
}
\author[DeGraf et al.]  
{\parbox{20cm}{Colin DeGraf$^{1,2}$, Debora Sijacki$^{1}$, Tiziana Di Matteo${^2}$, Kelly Holley-Bockelmann$^{3}$, \\ Greg Snyder$^{4}$, and Volker Springel$^{5}$} \vspace*{0.3cm}\\
{$^1$} Institute of Astronomy and Kavli Institute for Cosmology, University of Cambridge, Madingley Road, Cambridge CB3 0HA, UK \\
{$^2$} McWilliams Center for Cosmology, Dept. of Physics,
Carnegie Mellon University, Pittsburgh, PA 15213, USA\\
{$^3$} Department of Physics and Astronomy, Vanderbilt University, Nashville, TN 37235, USA \\
{$^4$} Space Telescope Science Institute, 3700 San Martin Dr, Baltimore, MD 21218, USA\\
{$^5$} Max-Planck-Institut f\"{u}r Astrophysik,
  Karl-Schwarzschild-Stra\ss{}e 1, 85740 Garching
  bei M\"{u}nchen, Germany}

\def\simgt{\lower.5ex\hbox{$\; \buildrel > \over \sim \;$}}

\usepackage{color}

\usepackage{outlines}

\begin{document}

\date{Submitted to MNRAS}
\pubyear{2020}

\maketitle
\begin{abstract}
With projects like LISA and Pulsar Timing Arrays expected to detect gravitational waves from supermassive black hole mergers in the near future, it is key that we understand what we expect those detections to be, and maximize what we can learn from them. To address this, we study the mergers of supermassive black holes in the Illustris simulation, the overall rate of mergers, and the correlation between merging black holes and their host galaxies. We find these mergers occur in typical galaxies along the $M_{\rm{BH}}-M_*$ relation, and that between LISA and PTAs we expect to probe the full range of galaxy masses. As galaxy mergers can trigger star formation, we find that galaxies hosting low-mass black hole mergers tend to show a slight increase in star formation rates compared to a mass-matched sample. However, high-mass merger hosts have typical star formation rates, due to a combination of low gas fractions and powerful AGN feedback. Although minor black hole mergers do not correlate with disturbed morphologies, major mergers (especially at high-masses) tend to show morphological evidence of recent galaxy mergers which survive for $\sim$500 Myr. This is on the same scale as the infall/hardening time of merging black holes, suggesting that electromagnetic followups to gravitational wave signals may not be able to observe this correlation. We further find that incorporating a realistic timescale delay for the black hole mergers could shift the merger distribution toward higher-masses, decreasing the rate of LISA detections while increasing the rate of PTA detections. 

\end{abstract}

\section{Introduction}
\label{sec:intro}
It is well established that supermassive black holes (SMBHs) are found at the center of massive galaxies \citep{KormendyRichstone1995}, and that the black hole (BH) masses correlate strongly with host galaxy properties, suggesting an evolutionary link between galaxy formation and BH growth \citep[e.g.][]{Magorrian1998, Gebhardt2000, Graham2001, Ferrarese2002, Tremaine2002, HaringRix2004, Gultekin2009, McConnellMa2013, KormendyHo2013, Reines2015, Greene2016, Schutte2019}.  As such, a galaxy merger can provide the opportunity for SMBHs from the respective galaxies to migrate to the new galactic centre, become gravitationally bound, form a binary, and eventually merge \citep[for a seminal discussion, see][]{Begelman1980}.  

Over the past several years, gravitational wave (GW) signals from BH mergers have been detected by the LIGO-Virgo collaboration \citep{Abbott2016}, but so far these gravitational waves have been limited to those produced by mergers between stellar mass BHs.  The expected mergers between SMBHs at the centers of galaxies would produce much longer wavelength gravitational waves, which the current ground-based interferometers are not sensitive to.  However, the planned Laser Interferometer Space Antenna (LISA) space mission will be focused on lower-frequency GWs, and is aimed at detecting mergers between SMBHs, with sensitivity peaking at $\sim 10^4-10^7 {\mathrm M_\odot}$ \citep{AmaroSeoane2017}.  In addition, Pulsar Timing Array observations should be capable of detecting mergers at even higher masses, reaching BHs above $10^8 {\mathrm M_\odot}$ \citep{Verbiest2016, Desvignes2016, Reardon2016, Arzoumanian2018}.  The expected detections of GWs from these observations should provide a completely novel and powerful framework to measure BHs and their co-evolution with their host galaxies.

Gravitational wave detections of SMBH mergers can provide estimates for the rate at which SMBHs merge throughout the universe \citep[e.g.][]{Klein2016, Salcido2016, Kelley2017, Ricarte2018, Katz2020}, the role mergers play in setting the BH-galaxy scaling relations \citep[e.g.][]{VolonteriNatarajan2009, Simon2016, Shankar2016}, gas environment and accretion efficiencies of BHs \citep[e.g.][]{Kocsis2011, Barausse2014, Derdzinski2019}, and even the mechanism by which BH seeds form \citep[see, e.g.][]{Sesana2007, Ricarte2018, DeGraf2019}.  Beyond information about the BHs themselves, multimessenger studies which combine GW and electromagnetic observations have the potential to analyse the galaxies in which these mergers occur, directly linking supermassive BH mergers with galaxy properties of the host.  

Cosmological hydrodynamical simulations which incorporate both galaxy formation and SMBHs provide an excellent tool to investigate the connection between BH mergers detectable by gravitational waves and the properties and histories of the host galaxies in which they are found.  Current cosmological simulations \citep[e.g.][]{Vogelsberger2014a, Dubois2014, Schaller2015, Feng2016, Pillepich2018} are able to probe a wide range of spatial scales, with BHs ranging from $\sim10^4 -10^{10} {\mathrm M_\odot}$, combined with galaxies with resolved morphological structure \citep[e.g.][]{Snyder2015, Snyder2019}.  Since these simulations self-consistently model the growth of BHs and the co-evolution of the host galaxies in large numbers, they provide an ideal resource to make predictions for what we can expect from upcoming gravitational wave detections, associated electromagnetic followups, and how to interpret those observations from a theoretical framework.

In this paper, we use the Illustris simulation \citep{Nelson2015} to investigate the connection between BH mergers and the galaxies in which they take place, with a particular emphasis on the galaxy morphology of BH merger hosts.  Using the detailed BH data in the Illustris simulation, we are able to connect each BH merger to the galaxy in which it occurs, and thus generate a list of every BH merger each galaxy has hosted throughout its history.  This allows us to analyze how recent BH mergers correlate with galaxy properties (such as morphology and star formation rate), the timescale over which these correlations survive, and the implications this has for electromagnetic followup observations to GW detections.

The outline of this paper is as follows.  In Section~\ref{sec:method} we discuss the Illustris simulations used for this project and the BH models therein.  In Section~\ref{sec:hostgalaxies} we show the galaxies which are found to host recent BH mergers.  In Section~\ref{sec:hostevolution} we investigate how these galaxies evolve before and after the BH merger event.  In Section~\ref{sec:hostmorphologies} we characterize the morphologies of host galaxies, with an emphasis on observationally identifying recent galaxy mergers, and in Section~\ref{sec:sfr} we investigate the connection between recent BH mergers and the host galaxy star formation rates.  In Section~\ref{sec:mergerdelays} we estimate the impact of incorporating inspiral and binary hardening timescales into the BH merger model, particularly with regards to the rate and masses of merging BHs detectable via gravitational waves.  Finally, in Section~\ref{sec:conclusions} we summarize our conclusions.

\section{Method}
\label{sec:method}

In this work, we primarily use the Illustris\footnote{https://www.illustris-project.org} suite of simulations \citep{Nelson2015}, run using the moving-mesh code {\small AREPO} \citep{Springel2010}, focusing on the highest resolution simulation run with a periodic box $106.5\, {\rm Mpc}$ on a side.  This model has target gas cell mass $m_{\rm{gas}}=1.26 \times 10^6\, {\mathrm M_\odot}$ and dark matter particle mass $m_{\rm{DM}}=6.26 \times 10^6\, {\mathrm M_\odot}$, with a standard $\Lambda$CDM cosmology: $\Omega_{m,0}=0.2726$, $\Omega_{\Lambda,0}=0.7274$, $\Omega_{b,0}=0.0456$, $\sigma_8=0.809$, $n_s=0.963$, and $H_0=70.4 \, \rm{km}\, \rm{s}^{-1} \, \rm{Mpc}^{-1}$ \citep[consistent with][]{Hinshaw2013}.

The Illustris simulations include detailed models of the physics involved in galaxy formation and evolution, including primordial and metal-line cooling with a time-dependent UV background \citep{FaucherGiguere2009} including self-shielding \citep{Rahmati2013}; star formation with associated supernova feedback \citep{SpringelHernquist2003, Springel2005}; stellar evolution, gas recycling and metal enrichment \citep[see][]{Wiersma2009} with mass and metal loaded outflows \cite[see][]{OppenheimerDave2008, Okamoto2010, PuchweinSpringel2013}.  For a more complete description of the physics incorporated into these simulations, see \citet{Vogelsberger2014a, Genel2014, Sijacki2015}.

Of particular importance to this project is the BH model, which we briefly summarize here \citep[for more complete details, see][]{Sijacki2015}.  BHs are treated as collisionless sink particles, which are seeded at mass $M_{\rm{seed}}=10^5 h^{-1} {\mathrm M_\odot}$ into any halo with $M_{\rm{halo}} > 5 \times 10^{10}\, h^{-1} {\mathrm M_\odot}$ which does not already contain a BH, loosely motivated by the direct collapse model \citep[see e.g.][]{HaehneltRees1993, LoebRasio1994, BrommLoeb2003, Begelman2006, ReganHaehnelt2009}, but intended to remain broadly consistent with lighter seed formation models followed by relatively efficient mass growth.  After seeding, Illustris grows BHs by gas accretion modelled by a Bondi-Hoyle-like rate \citep{BondiHoyle1944, Bondi1952}.  During gas accretion, three modes of BH feedback are included which deposit energy onto the surrounding material or alter its cooling rate (“quasar”, “radio”, and “radiative”, depending on the accretion efficiency).  BHs which come within each other's smoothing lengths merge together into a single BH.  The simulation saves the exact time and masses of the merger, giving us precise information for each merger which occurs within the simulation. In addition to the merger events themselves, this provides us with a complete merger tree for every BH, which we use to extract the most recent merger events which occur in each galaxy.  We note that Illustris does not incorporate a hardening time for the binary, and instead assumes instantaneous coalescence once the BHs are within smoothing lengths of each other, the implications of which we investigate in Section~\ref{sec:mergerdelays}.  

In addition to the BH model, Illustris identifies dark matter haloes with a friends-of-friends (FOF) halo finder \citep{Davis1985}, and assigns baryonic matter to the same group as the nearest dark matter particle.  After the FOF finder, the {\small SUBFIND} algorithm \citep{Springel2001, Nelson2015} produces a catalog of gravitationally self-bound substructures, and the mergers between these subhalos are tracked using the SubLink catalog \citep{Rodriguez-Gomez2015}.  The SubLink algorithm sums a merit function for every particle found in common between subhalos in different snapshots, and identifies the subhalo descendant as the subhalo with the highest summed merit score.  Thus we have a complete list of galaxy properties (from {\small SUBFIND}) and a complete merger tree for all resolved subhaloes (SubLink) from which we can connect BHs and their host galaxies, as well as the complete merger history for each such host.

\section{Galaxy hosts of supermassive black hole mergers}
\label{sec:mergerhosts}

\begin{figure}
    \centering
    \includegraphics[width=0.48\textwidth]{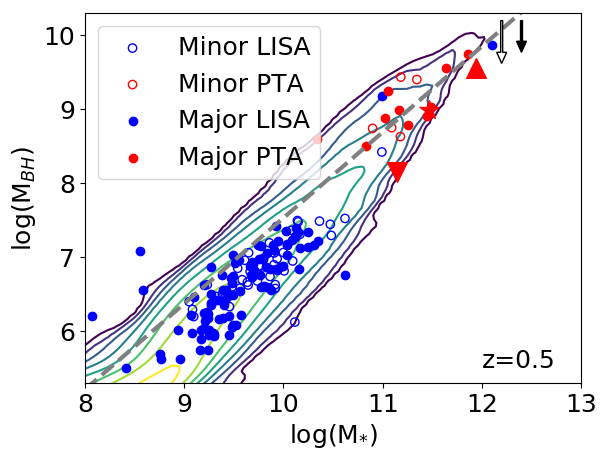}
    \caption{$M_{\rm{BH}}-M_*$ relation at $z=0.5$. The contours show the overall distribution of galaxies in the Illustris simulation (the outermost contour encompasses $\sim 99.8 \%$ of galaxies, while the innermost encompasses $\sim 16 \%$), with a comparison to the local relation \citep[][dashed grey line]{KormendyHo2013}.  The coloured points are the subsample of galaxies which have hosted a BH merger in the past 300 Myr, divided into major ($M_2/M_1 > 0.25$, filled circles) vs. minor ($0.1 < M_2/M_1 < 0.25$, open circles) BH mergers, and LISA-type ($M_{\rm{chirp}} < 10^7\, {\mathrm M_\odot}$, blue) vs PTA-type ($M_{\rm{chirp}} > 10^8 \: {\mathrm M_\odot}$) mergers.  $M_{\rm{BH}}$ here is defined as the mass of the central (i.e. most massive) BH in the galaxy.  The solid and open arrows show the typical difference between central BH mass and chirp mass of the merger.  The three larger symbols (star, upward triangle, downward triangle) correspond to specific targets investigated in more detail in Figs.~\ref{fig:evo1}-\ref{fig:evo3}.  }
    \label{fig:scalingrelation}
\end{figure}

\subsection{Host galaxies}
\label{sec:hostgalaxies}

We start our analysis by looking at the galaxies which host SMBH mergers, in particular focusing on the scaling relation between BH mass ($M_{\rm{BH}}$) and galaxy stellar mass ($M_*$).  In Fig.~\ref{fig:scalingrelation} we show the relation at $z=0.5$, which approximately matches the observed local scaling relation \citep[for more analysis of the scaling relation in Illustris, see][]{Sijacki2015}.  Here we highlight the galaxies which have, in the past 300 Myr, hosted mergers at scales expected to be detectable by Pulsar Timing Arrays (PTA) or LISA.  In blue we show LISA-detectable mergers, defined as BH mergers with chirp mass $M_{\rm{chirp}} \coloneqq (M_1 M_2)^{3/5}/(M_1+M_2)^{1/5} < 10^7\, {\mathrm M_\odot}$ (with no minimum since the LISA sensitivity extends below the seed mass of BHs in the simulation), and in red the PTA-detectable mergers defined as those with $M_{\rm{chirp}} > 10^8\, {\mathrm M_\odot}$.  We also subdivide the mergers based on the ratio between the larger ($M_1$) and smaller ($M_2$) masses involved in the merger, with minor BH mergers defined as BH mass ratios of $0.1 < M_2/M_1 < 0.25$, and major BH mergers defined by $M_2/M_1 > 0.25$.  As expected, the high-mass PTA-detectable mergers are found exclusively in high-mass galaxies, while lower-mass LISA-detectable mergers tend to be found in lower mass galaxies. 

We note there are three outliers: LISA-type mergers which occur in high-mass galaxies, with $M_{\rm{BH}} >> 10^8\, {\mathrm M_\odot}$, well above our mass scale of $M_{\rm{chirp}} < 10^7\, {\mathrm M_\odot}$.  These represent massive galaxies with high-mass central BHs (i.e. high $M_{\rm{BH}}$) which host a merger between two low-mass satellite BHs (i.e. low BH merger masses, and thus low $M_{\rm{chirp}}$).  This is a rare occurrence, consisting of only $\sim 2\%$ of LISA-type mergers within Illustris\footnote{There is a possibility of having numerical issues with halo finders near galaxy mergers, which could potentially result in spurious generation of BH seeds. However, this is rare in simulations which seed based on halos (rather than subhalos) such as Illustris, and we have explicitly confirmed that the outlying LISA-type mergers are indeed satellite BHs with low masses compared to the host galaxy and not due to spurious seeding.}.
For the remainder of cases, however, we find that $M_{\rm{chirp}}$ and $M_{\rm{BH}}$ are similar, with a mean ${\rm log}(M_{\rm{BH}}/M_{\rm{chirp}}) = 0.5\pm0.04$ ($0.6\pm0.04$) for major (minor) BH merger hosts (characterized by filled and open arrows in Fig.~\ref{fig:scalingrelation}, respectively).   However, Illustris only seeds each galaxy with a single BH, and the minimum masses involved are limited by the seed mass ($M_{\rm{seed}}=10^5 \: h^{-1} \: {\mathrm M_\odot}$); thus the frequency of outlier mergers (of BHs undermassive relative to their host galaxy) should be expected to be higher than predicted here.  Overall, this suggests that galaxies hosting BH mergers have typical $M_{\rm{BH}}/M_*$ ratios and generally have $M_{\rm{chirp}}$ within $\sim 0.5$ dex of $M_{\rm{BH,cen}}$.  Furthermore, while both LISA and PTAs are limited in the mass ranges they can probe, between the two types of detectors we can expect to probe typical host galaxies across a wide range of masses (covering the entire range resolved by the Illustris simulation).

\begin{figure}
    \centering
    \includegraphics[width=8.5cm]{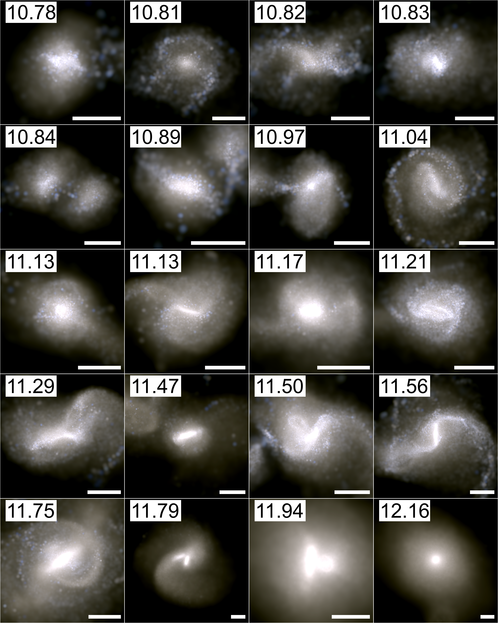}
    \caption{Montage \citep[images from Illustris public release, see][]{Nelson2015} of $z=0.5$ galaxies that hosted a BH merger with $M_2 > 10^7\, {\mathrm M_\odot}$ (thus roughly PTA-type mergers) and $M_2/M_1 > 0.25$ in the prior 300 Myr, showing a mix of disturbed and relatively relaxed morphologies. The galaxies are ordered by stellar mass, from least massive in the upper left to most massive in the lower right, with log($M_*/M_\odot$) of the galaxy and a 20 ckpc scalebar provided in each panel. Here we see the majority of hosts show clearly disturbed morphologies, including asymmetries, tidal tails, light shells and dual nuclei, indicative of recent merger activity.}
    \label{fig:montage05}
\end{figure}

However, we find that galaxies hosting supermassive black hole mergers are not morphologically typical, although their masses are.  To demonstrate this qualitatively, in Fig.~\ref{fig:montage05} we show mock galaxy images in observed-frame James Webb Space Telescope bands (neglecting the effect of dust) at $z=0.5$ for hosts of recent major, massive BH mergers (ordered by galaxy stellar mass, from least massive in the upper left to most massive in the lower right).  Here we define a recent merger as having occurred in the previous $300$~Myr, a major BH merger as one with a mass ratio $M_2/M_1 > 0.25$, and a massive BH merger as one in which $M_2 > 10^7\, {\mathrm M_\odot}$.   We note that this definition of a massive BH merger is less stringent than the PTA constraint used in Fig.~\ref{fig:scalingrelation}, to provide us with a more reasonable sample size for the following statistical analysis, but this change has no qualitative impact on Fig.~\ref{fig:scalingrelation}.  

\begin{figure}
    \centering
    \includegraphics[width=8.5cm]{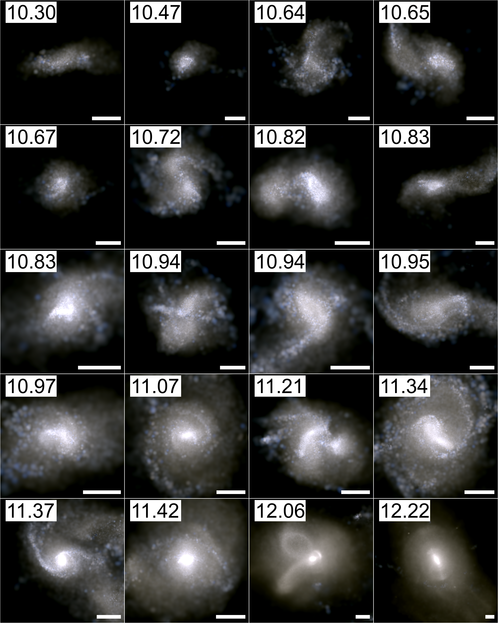}
    \caption{As in Fig.~\ref{fig:montage05}, but at $z=1$.  As in Fig.~\ref{fig:montage05}, we see clear evidence of disturbed morphologies, and also note that, as expected, the galaxies are noticeably bluer than those at $z=0.5$.}
    \label{fig:montage1}
\end{figure}

\begin{figure}
    \centering
    \includegraphics[width=8.5cm]{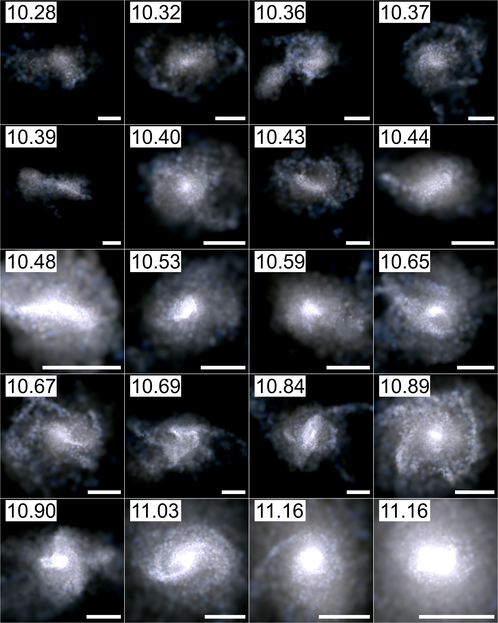}
    \caption{As in Fig.~\ref{fig:montage05}, but at $z=2$.  As in Figs.~\ref{fig:montage05}-\ref{fig:montage1} we see disturbed morphologies, but less universally so across the full sample, with some galaxies displaying unperturbed disc-like structure.}  
    \label{fig:montage2}
\end{figure}

Fig.~\ref{fig:montage05} shows that from a simple visual analysis, many of the galaxies hosting massive BH major mergers exhibit disturbed morphologies, providing evidence for galaxy mergers in the form of dual nuclei, secondary satellite galaxies, shells, or tidal features.  In Figs.~\ref{fig:montage1} and \ref{fig:montage2}, we show the galaxies hosting massive major BH mergers at $z=1$ and $z=2$, respectively.  As in the case of the general galaxy population, we see that high-redshift merger hosts tend to be much bluer than those at low-$z$, and the low-mass galaxies (upper left panels) tend to be bluer than the higher mass galaxies (lower right panels).  The higher redshift images show disturbed morphologies, but these disturbances are less ubiquitous than at low-$z$ (in particular regarding dual nuclei and distinct satellite galaxies).  Thus we find that galaxies which host massive major BH mergers can show morphological evidence for a galaxy merger, with the strongest evidence occurring at lower redshifts (which we investigate quantitatively in Section~\ref{sec:hostmorphologies}).

\subsection{Host evolution}
\label{sec:hostevolution}

\begin{figure*}
    \centering
    \includegraphics[width=18cm]{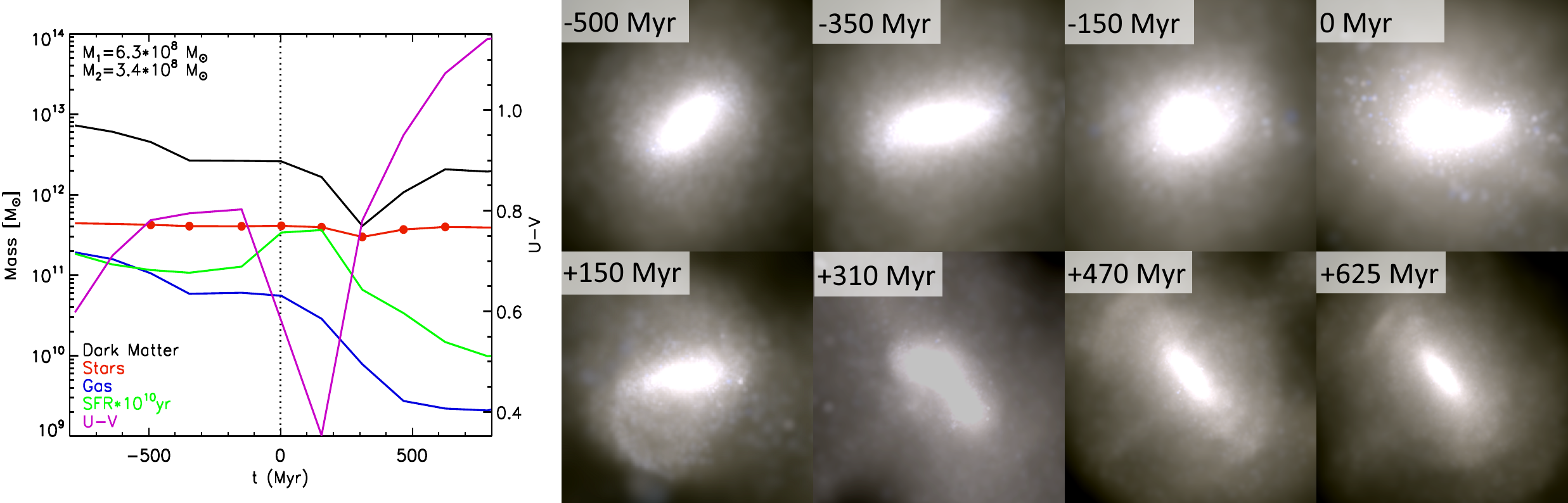}
    \caption{The evolutionary history of a galaxy hosting a high-mass major BH merger.  \textit{Left panel:} $M_{\rm{DM}}, M_*, M_{\rm gas}$, SFR, and U-V colour for the galaxy as a function of time since the BH merger.  \textit{Right panels:} Stellar composite images from sequential snapshots, corresponding to the circles marking snapshots in the left panel.  The time since the BH merger is marked in each panel, with the BH merging at the galaxy centre at time $t=0$~Myr. Initially we see a well defined nucleus, followed by a disturbed nucleus as a secondary galaxy has its first passage, followed by a defined nucleus with light shells showing evidence of the recent merger.  Note that this galaxy corresponds to the star symbol used in Figs.~\ref{fig:scalingrelation}, \ref{fig:gini_m20}, \ref{fig:s_vs_mergetime}, and \ref{fig:ssfr_mtot}.}
    \label{fig:evo1}
\end{figure*}

\begin{figure*}
    \centering
    \includegraphics[width=18cm]{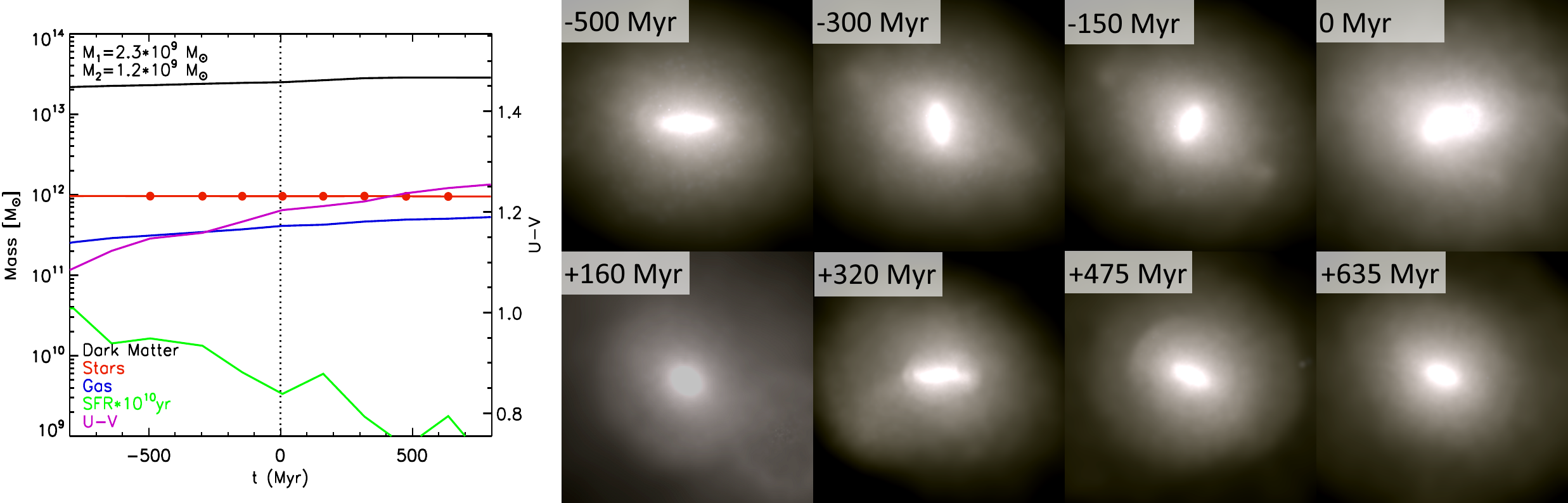}
    \caption{As in Fig.~\ref{fig:evo1}, but for another galaxy. Here we see less disturbance in the nucleus, but nonetheless long-lived, well defined light shells.  This galaxy corresponds to the upward pointing triangle symbol in Figs.~\ref{fig:scalingrelation}, \ref{fig:gini_m20}, \ref{fig:s_vs_mergetime}, and \ref{fig:ssfr_mtot}.}
    \label{fig:evo2}
\end{figure*}

\begin{figure*}
    \centering
    \includegraphics[width=18cm]{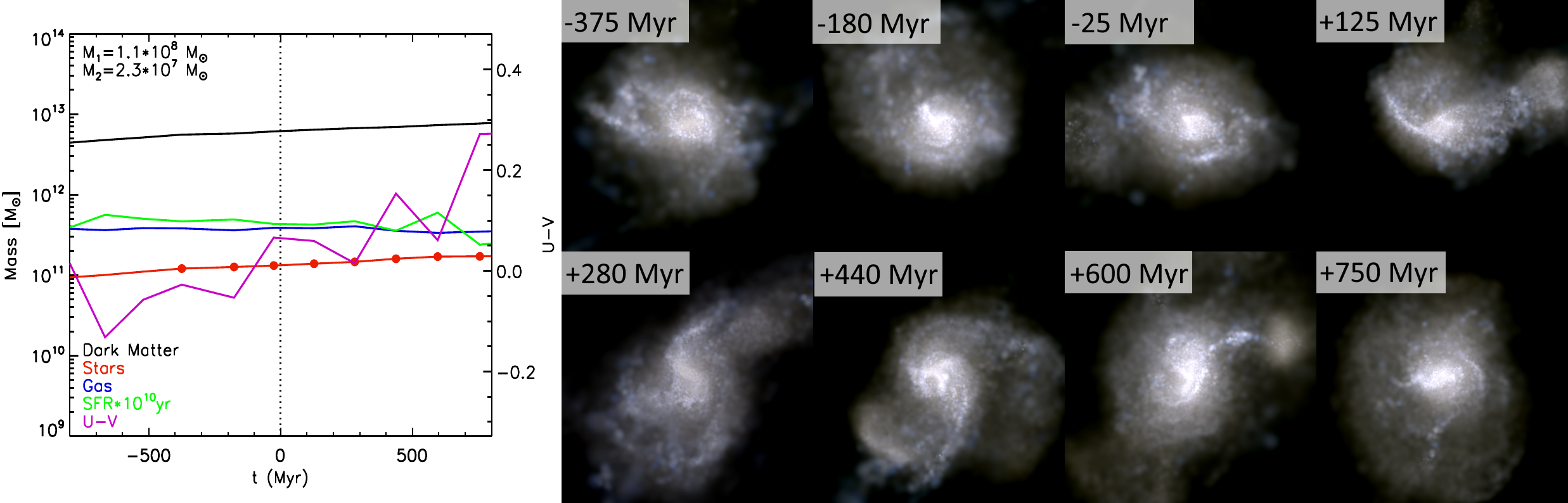}
    \caption{As in Fig.~\ref{fig:evo1}, but for a particularly gas rich galaxy.  Here we do not see any light shells, but do observe a highly disturbed morphology caused by the galaxy merger.  This galaxy corresponds to the downward pointing triangle symbol in Figs.~\ref{fig:scalingrelation}, \ref{fig:gini_m20}, \ref{fig:s_vs_mergetime}, and \ref{fig:ssfr_mtot}.}
    \label{fig:evo3}
\end{figure*}

\begin{figure*}
    \centering
    \includegraphics[width=18cm]{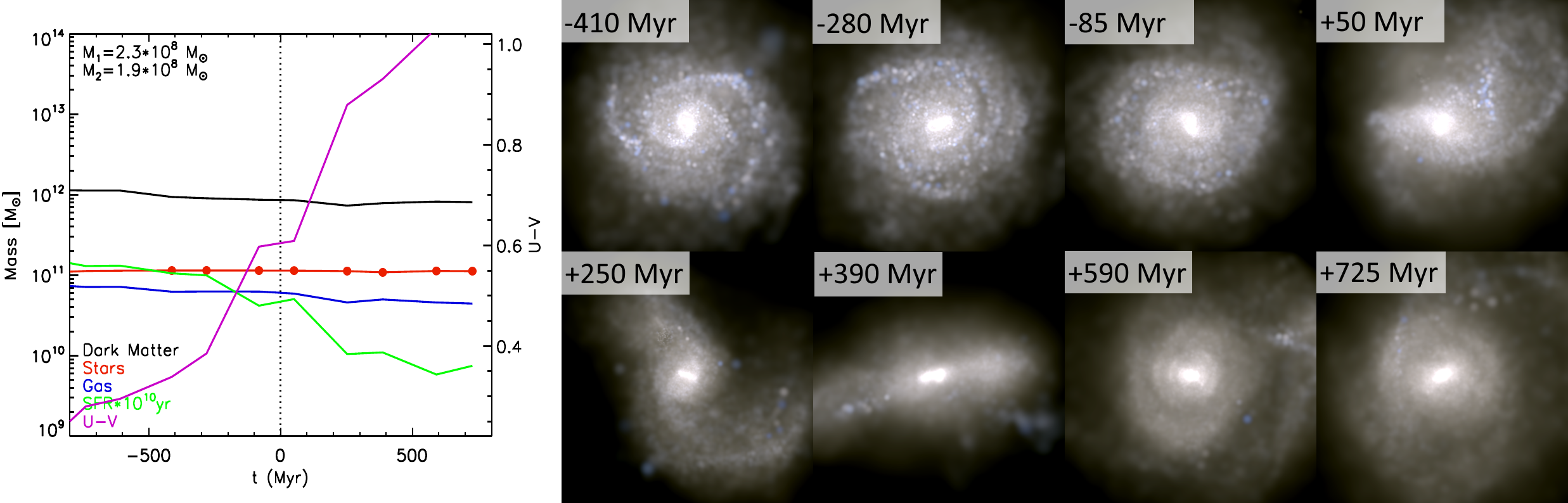}
    \caption{As in Fig.~\ref{fig:evo1}, but for a galaxy from the TNG100 simulation at $z=0$.  As in Fig.~\ref{fig:evo3} we see a clearly disturbed morphology corresponding to the BH merger, and we also witness significant reddening which occurs following the merger.}
    \label{fig:evotng}
\end{figure*}

To better understand how galaxies hosting supermassive black hole mergers will evolve with time, in Figs.~\ref{fig:evo1}-\ref{fig:evotng} we track individual galaxies, over $\sim$ 1 Gyr, spanning both before and after the merger event.  In each figure we show the evolution of host galaxy properties ($M_{\rm DM}$, $M_{*}$, $M_{\rm gas}$, star formation rate, and U-V colour) in the leftmost panel, and the galaxy morphology at each snapshot in the right hand panels.  The morphology images are in sequential order starting at the upper left panel, with each panel labelled with the time relative to the BH merger. 

In Fig.~\ref{fig:evo1} we show the evolutionary history of a galaxy which hosts a merger between black holes with mass $M_1=6.3 \times 10^8\, {\mathrm M_\odot}$ and $M_2=3.4 \times 10^8\, {\mathrm M_\odot}$.  Pre-merger we see a well-defined stellar nucleus.  Near the BH merger event we see a disturbed nucleus, produced as the satellite galaxy has a first passage and when the post-merger galaxy settles down.  At later times, we see clear light shells as remnants of the galaxy merger, surviving for an extended period ($\sim$500 Myr).  We see an increase in star formation, with a corresponding change to bluer colour, near the merger time, as the galaxy merger triggers a burst of star formation \citep{2005ApJ...620L..79S}.  This is followed by a significant drop in star formation (SF) and gas mass ($M_{\rm{gas}}$ decreases by $\sim$1.5 orders of magnitude in the $500\,{\rm Myr}$ following the merger), and a corresponding reddening of the galaxy as the SF falls off.    

In Fig.~\ref{fig:evo2} we show the history of a galaxy hosting a more massive BH merger ($M_1=2.3\times 10^9\, {\mathrm M_\odot}$ and $M_2=1.2\times 10^9\, {\mathrm M_\odot}$).  Here we again see a decreasing star formation rate (SFR) and mildly increasing U-V colour, but with an overall SFR that is much lower, and the gas supply remains relatively high throughout the merger period.  Morphologically, at the time nearest the merger we see evidence for a dual nucleus, and again there are light shells which survive long past the merger event (though fainter than in Fig.~\ref{fig:evo1}).  

In Fig.~\ref{fig:evo3} we consider the history of a galaxy hosting a less massive, minor BH merger ($M_1=1.1 \times 10^8\, {\mathrm M_\odot}$ and $M_2=2.3 \times 10^7\, {\mathrm M_\odot}$, note that this is near the threshold for a major BH merger), which occurs in a very gas-rich galaxy.  Here we again see a reddening of the galaxy; also the morphology is highly disturbed immediately following the BH merger, with tidal features visible for several hundred Myr.  

We also consider the impact of the black hole model used in the simulation.  In particular, the Illustris simulation has previously been found to have Active Galactic Nucleus (AGN) feedback which could too easily heat and dilute the central gas in galaxy groups, whereas IllustrisTNG uses a new feedback model to resolve this issue \citep{Weinberger2017}.  However, the effect is strongest among highest-mass galaxies, so we do not expect significant qualitative impact.  Nonetheless, to test the sensitivity to the assumptions made regarding black hole modeling in the simulation, we also show an evolutionary history from the TNG100 model of the IllustrisTNG\footnote{https://www.tng-project.org} suite of simulations \citep{Marinacci2018, Springel2018, Nelson2018, Pillepich2018b, Naiman2018}, at $z=0$ in Fig.~\ref{fig:evotng}.  
Prior to the merger in this example, we see a well defined spiral galaxy, which is noticeably disturbed near the time of the BH merger.  As in Figs.~\ref{fig:evo1} and \ref{fig:evo2}, we find a strong decrease in SF post-merger, with a correspondingly significant increase in U-V colour.  The similarity between these histories from Illustris and TNG100 suggests that the qualitative results found here do not exhibit a strong sensitivity to the black hole model of the simulation.
Overall, this subsample suggests that galaxies hosting a massive BH merger generally show morphological evidence of a merger, and tend to redden and decrease in SF during the post-merger phase, which we quantify for the full population in Sections~\ref{sec:hostmorphologies} and \ref{sec:sfr}.

\subsection{Host morphologies}
\label{sec:hostmorphologies}

Having seen the morphologies of galaxies hosting major massive BH mergers in Figs.~\ref{fig:montage05}-\ref{fig:montage2} and individual evolutions of several sample galaxies in Figs.~\ref{fig:evo1}-\ref{fig:evotng}, we now consider the morphology statistics for the full host sample relative to the full galaxy population.  

\begin{figure}
    \centering
    \includegraphics[width=0.49\textwidth]{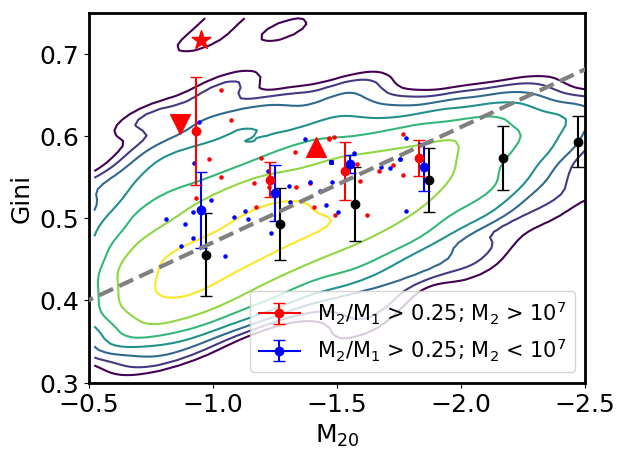}
    \caption{The Gini - $M_{20}$ relation at $z=0.5$ for galaxies with $M_* \gtrapprox 10^{9.7} {\mathrm M_\odot}$ (contours, from the outermost encompassing $\sim 99.7 \%$ of galaxies to the innermost encompassing $\sim 30 \%$), for galaxies which have hosted a BH merger in the past 300 Myr with BH mass ratio $M_2/M_1 > 0.1$ (red points) and $M_2/M_1 > 0.1 ; M_2 > 10^7\, {\mathrm M_\odot}$ (blue points), and error bars which show the mean and standard deviation for each population.  Galaxies with a recent massive major BH merger tend to have higher Gini than the general galaxy sample, corresponding to what we would expect a galaxy merger to exhibit.  Low-mass major BH mergers have a slightly higher Gini than the full galaxy sample, but the trend is much weaker and only at the order of $\sim 1 \sigma$.  The star, upward triangle, and downward triangle correspond to the individual histories plotted in Fig.~\ref{fig:evo1}, \ref{fig:evo2}, and \ref{fig:evo3}, respectively.}
    \label{fig:gini_m20}
\end{figure}

In order to quantify each galaxy, we use Gini \citep[a measure of how evenly the galaxy flux is distributed, from $1$ if all flux is in a single pixel to $0$ for a uniform distribution across all pixels; see][for more details]{Lotz2004} and $M_{20}$ (defined as the second-order moment of the brightest 20\% of pixels) parameters as simple measures of morphology. In Fig.~\ref{fig:gini_m20} we show the Gini-$M_{20}$ relation for galaxies at $z=0.5$.  The coloured contours show the distribution of all galaxies with $M_* > 10^{9.7}\, {\mathrm M_\odot}$ in Illustris.  ``Normal'' non-merging galaxies will tend to lie toward the bottom right of the panel, while merging galaxies will be more likely to lie in the upper left portion \citep{Lotz2004}; merging galaxy are more likely to have a high Gini coefficient (i.e. stellar light concentrated in a smaller fraction of total pixels) and a high $M_{20}$ (i.e. the brightest 20\% of pixels are less centrally concentrated) compared to a non-merging galaxy.  Here we plot a dashed line roughly differentiating between morphologies suggesting a merger and those suggesting a quiescent galaxy \citep{Snyder2019}.  The datapoints show the location of individual galaxies hosting major massive BH mergers (red) and major low-mass BH mergers (blue) in the past 300 Myr, and the $M_{20}$-binned mean Gini values (coloured errorbars) compared to the mean of the full population (black errorbars).  Here we find that recent major BH mergers tend to lie $\sim 1 \sigma$ above the general population, and $>2 \sigma$ above for high-mass mergers in high-$M_{20}$ galaxies, consistent with the earlier qualitative results.

To better characterize this difference, we use the merger statistic $S$ \citep[see][]{Snyder2019}, defined as:
\begin{equation}
S(G,M_{20}) = 0.139 M_{20} + 0.990G - 0.327\,,
\label{eq:sstat}
\end{equation}
(where G is the Gini coefficient) which provides a single parameter to quantify how merger-like the galaxy morphology is, with a positive $S(G,M_{20})$ corresponding to a merger-like galaxy and negative $S(G,M_{20})$ being non merger-like.  

\begin{figure}
    \centering
    \includegraphics[width=0.49\textwidth]{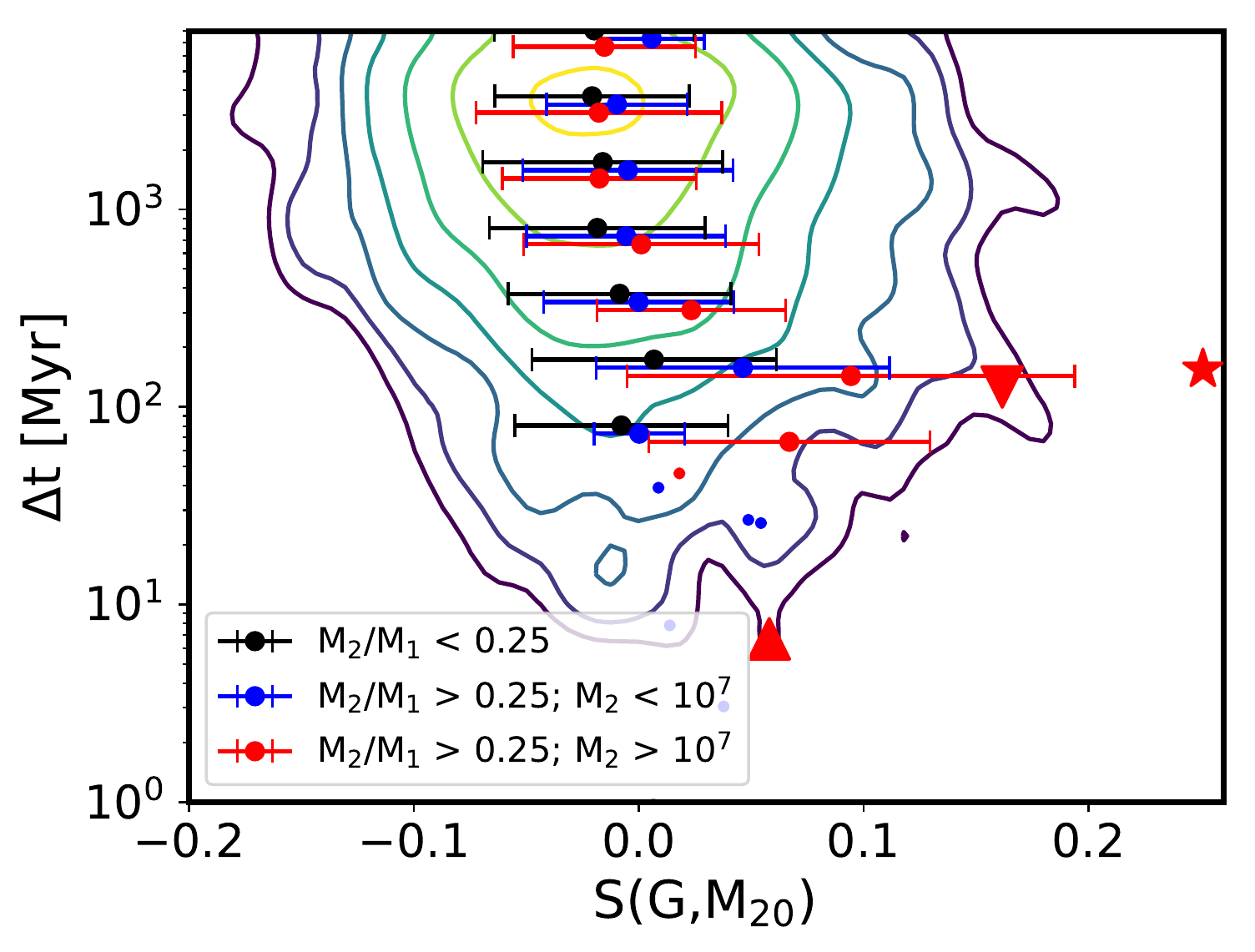}
    \caption{Merger statistic $S$ (see Equation~\ref{eq:sstat}) for $z=0.5$ galaxies as a function of time since the most recent BH merger within the galaxy (contours, ranging from $\sim 80 \%$ of galaxies to $\sim 11 \%$).  Also shown are the mean and standard deviation of $S$ statistics binned by time since the most recent major BH merger ($M_2/M_1 > 0.25$, blue points), since the most recent massive major BH merger ($M_2/M_1 > 0.25; M_2 > 10^7\, {\mathrm M_\odot}$, red points), and since the most recent non-major BH merger ($M_2/M_1 < 0.25$, black points).} 
    \label{fig:s_vs_mergetime}
\end{figure}

In Fig.~\ref{fig:s_vs_mergetime} we characterize the sensitivity of the galaxy morphology on time since the BH merger, with the contours showing the time since the most recent BH merger ($\Delta t$) against $S(G,M_{20})$.  The error bars show the mean and standard deviation of $S(G,M_{20})$ binned by time since the most recent non-major BH merger (black), low-mass major BH merger (blue), and high-mass major BH merger (red).  For the binned data, we consider the time of the most recent merger, but do not include galaxies if the most recent merger is more ``significant'' than specified for each population (i.e.~red points calculate the time since the most recent massive major BH merger; blue points calculate the time since the most recent low-mass major BH merger, but ignore galaxies where there was a more recent high-mass major BH merger).

For $\Delta t > 1$~Gyr, we find no difference between the merger populations, and for $\Delta t > 500$~Myr, there are only very minor differences ($<< 1\sigma$).  However, for $\Delta t < 300$~Myr, we find that massive major BH merger hosts have a significantly higher $S(G,M_{20})$, while low-mass major BH merger hosts show a moderately higher $S(G,M_{20})$, though at $<1 \sigma$ significance.

To further demonstrate this correlation between recent mergers and host morphology, in Fig.~\ref{fig:1d_histograms} we show the $z=0.5$ probability (top) and cumulative (bottom) distribution functions of $S(G,M_{20})$ for all galaxies (black) and for galaxies which, in the past 300 Myr, have hosted a low-mass major BH merger (blue) or high-mass major BH merger (red).  As in Figs.~\ref{fig:gini_m20} and \ref{fig:s_vs_mergetime}, we see that low-mass major BH mergers lie $\sim 1 \sigma$ above the general galaxy sample, while the high-mass major BH mergers are $\sim 1.5 \sigma$ higher, confirming that hosts of major BH mergers are significantly morphologically disturbed.  

In addition, we show the equivalent distributions at $z=1$ (middle panels) and $z=2$ (right panels).  At $z=1$ we find that low-mass major BH mergers are no longer significantly different from the general population, while the high-mass major BH mergers are still $\sim 1 \sigma$ above the general population.  By $z=2$, we see that the galaxies typically have the same $S(G,M_{20})$ distribution regardless of whether they have hosted a recent BH merger or not.  Thus, we conclude that recent BH mergers correlate with merger morphologies at low-redshift, though above $z \sim 1$ there is no longer any significant correlation.  We confirm the statistical significance of these claims using a Kolmogorov-Smirnov test, finding that both major BH merger populations at $z=0.5$, and the high-mass major BH merger population at $z=1$, each have a $<0.2\%$ likelihood of being drawn from the same distribution as the full galaxy sample (Table~\ref{tab:kstest}).  We also checked other $\Delta t$, and the only longer timescale for which a merger signal survives is for massive major BH merger hosts at $z=0.5$, in which the signal survives for $\sim 500$~Myr (the other cases differ from the general population by $<0.5\sigma$).  We note that this survival time of $\sim$ 500 Myr for a morphological signal is on the order of the galaxies' dynamical time, consistent with a picture in which the galaxies gradually return to a relaxed state.

\begin{table*}
    \centering
    \begin{tabular}{c|c|c|c}
        \hline
          & Time since BH merger [Myr] & Low-mass major BH mergers & High-mass major BH mergers \\
        \hline
         $z=2.0$  & 0-300 & $0.70$ & $0.64$ \\ 
         $z=1.0$  & 0-300 &  $0.27$ & \fbox{$1.8 \times 10^{-3}$} \\
         $z=0.5$ & 0-300 & \fbox{$3.6 \times 10^{-4}$} & \fbox{$5.0 \times 10^{-6}$} \\
          & 300-500 & $0.45$ & \fbox{$3.4 \times 10^{-4}$} \\
          & 500-1000 & $0.22$ & $0.16$ \\
         \hline
    \end{tabular}
    \caption{The KS-test likelihood that the target sample (high- or low-mass major BH merger hosts) comes from the same $S(G,M_{20})$ distribution as the full galaxy sample (see Fig.~\ref{fig:1d_histograms}).  At $z=0.5$, low mass major BH merger hosts have a signal which survives $\sim 300$~Myr while the signal in high mass major BH merger hosts survives for $\sim 500$~Myr.  At higher redshifts, the signal is weaker; only the high-mass hosts have a signal at $z=1$, which survives for $\sim 300$~Myr.}
    \label{tab:kstest}
\end{table*}

We note that massive BH mergers tend to occur in high-mass galaxies (as shown in Fig.~\ref{fig:scalingrelation}), which could potentially contribute to the difference between the general galaxy population and the massive major BH merger hosts.  As such, we compute a distribution of the general galaxy population which is a mass-matched equivalent to the merger host sample.  
The mass-matched equivalent is found by weighting each galaxy in the full population sample by the ratio between a Gaussian best-fit to the target sample's $M_{\rm tot}$ distribution (neglecting the 5\% furthest outliers in each direction) and the full galaxy population's $M_{\rm tot}$ distribution.  The result is a weighted PDF of the full galaxy sample but with an equivalent galaxy-mass distribution to the target sample. We plot the weighted distributions for these mass-matched samples as dashed lines in Fig.~\ref{fig:1d_histograms}.  At all redshifts, the mass-matched samples (dashed lines) match the full galaxy population (solid black), which confirms that the $S(G,M_{20})$ distributions are not dependent on galaxy mass, and the significant increase among BH merger hosts is genuinely due to merging.

\subsection{Star formation rates}
\label{sec:sfr}

\begin{figure}
    \centering
    \includegraphics[width=0.49\textwidth]{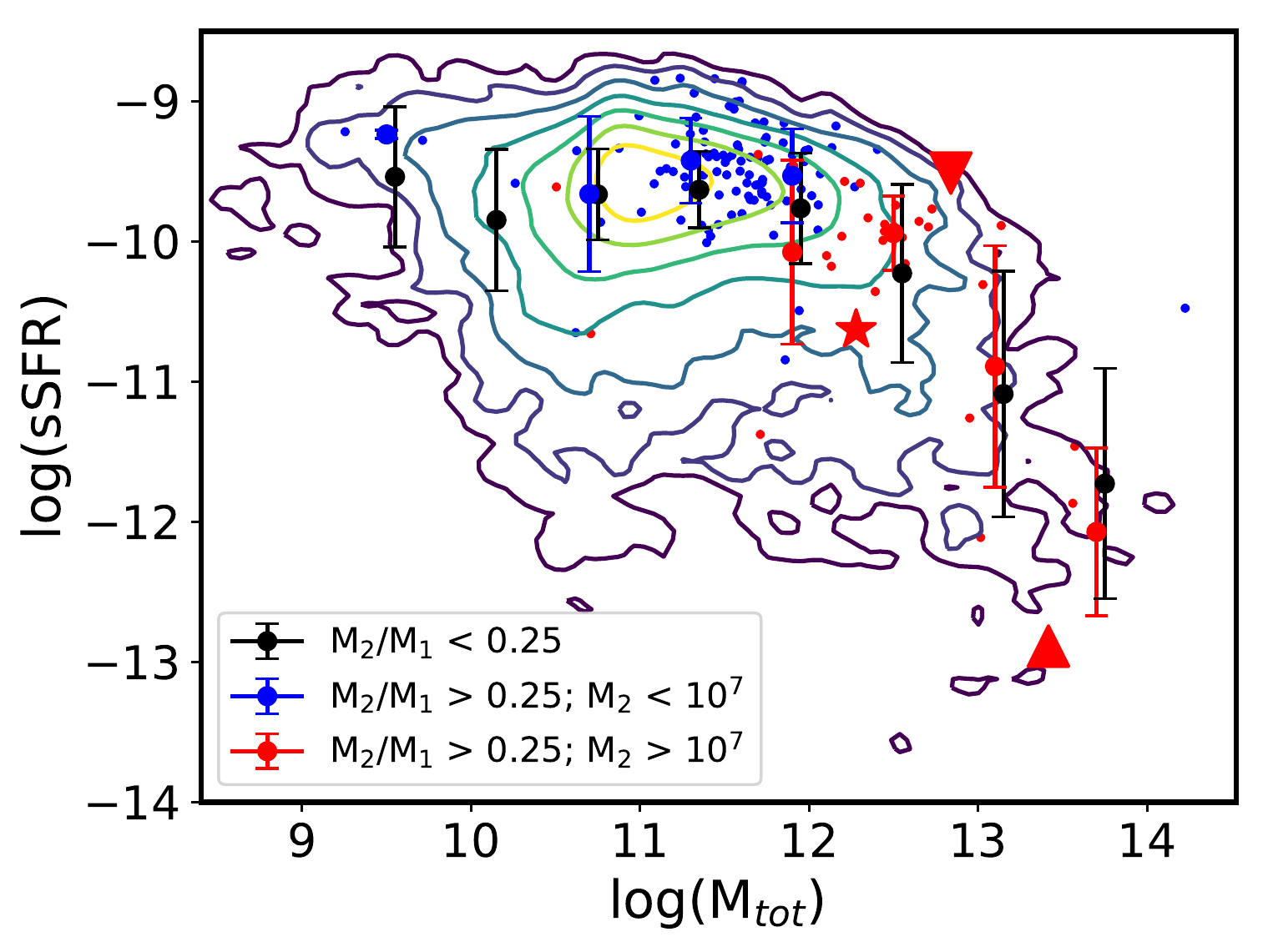}
    \caption{Specific star formation rate vs. total galaxy mass for the full galaxy sample (contours, with the outermost encompassing $\sim 99.7 \%$ of galaxies, and the innermost encompassing $\sim 46 \%$ of galaxies) and high-mass (red) and low-mass (blue) major BH merger hosts.  We find that high-mass major BH mergers tend toward high-mass hosts and thus lower sSFR, but are fully consistent with a mass-matched sample from the full galaxy population.}
    \label{fig:ssfr_mtot}
\end{figure}

Given the potential for galaxy mergers to influence typical host star formation, and the substantial post-merger decrease in SFR seen in the individual histories in Figs.~\ref{fig:evo1}, \ref{fig:evo2}, and \ref{fig:evotng}, we also look for correlations between BH mergers and host star formation rates.  In Fig.~\ref{fig:ssfr_mtot}, we plot the specific star formation rate (sSFR$\coloneqq$SFR/$M_*$) vs. total galaxy mass at $z=0.5$.  As expected, we see that the sSFR tends to drop among high-mass galaxies (contours and black points), since high-mass galaxies tend to be less star-forming.  We also plot the populations for the high- (red) and low- (blue) mass major BH merger hosts, together with the $M_{\rm tot}$-binned mean and standard deviation (error bars). High-mass major BH mergers tend to have similar sSFR to other equivalent mass galaxies (red vs. black error bars).  However low-mass major BH mergers tend to have slightly higher sSFR for their masses (blue vs. black error bars), suggesting that those mergers tend to trigger some additional star formation.

\begin{figure*}
    \centering
    \includegraphics[width=16cm]{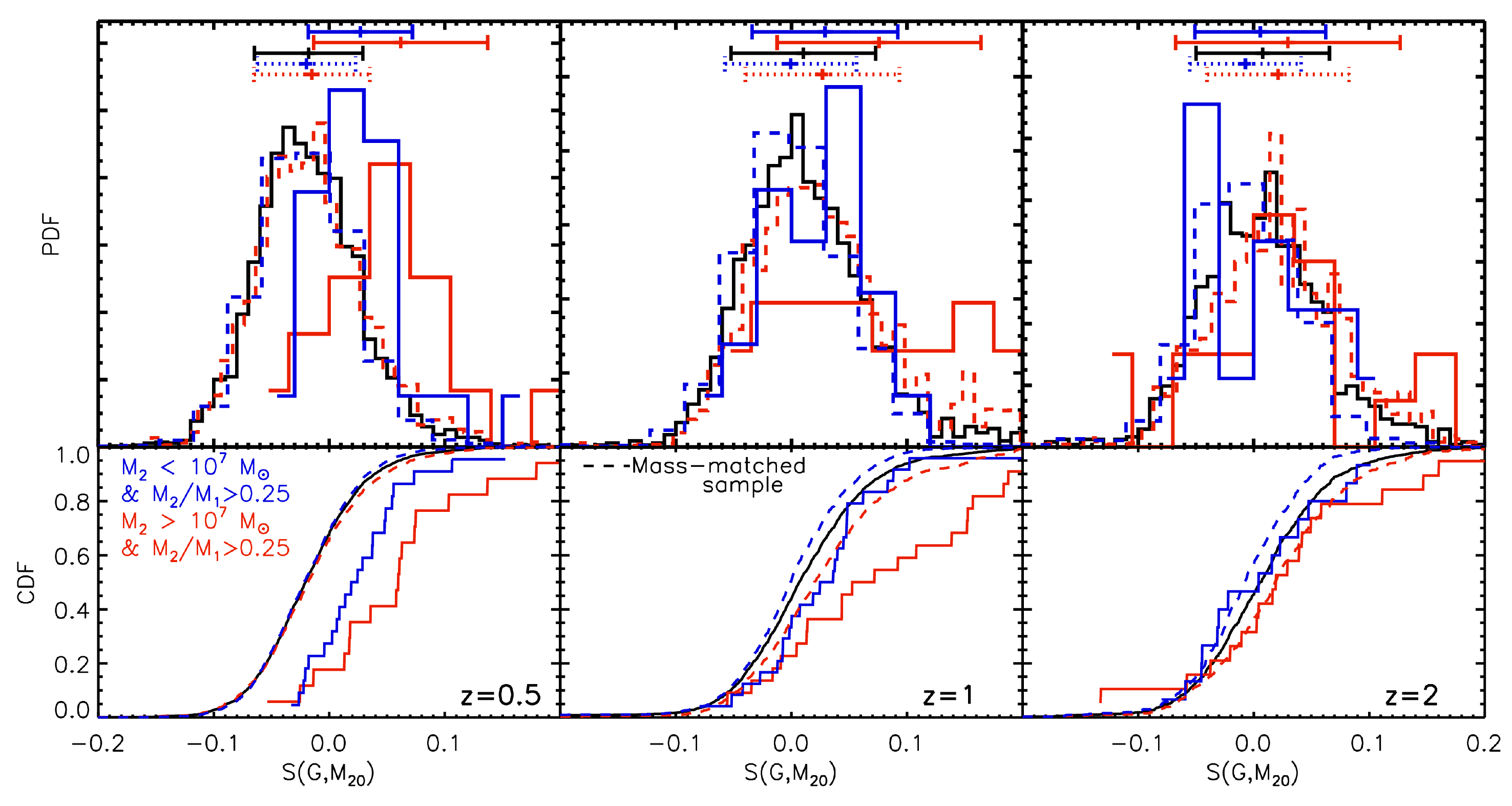}
    \caption{Probability distribution function (top panels) and cumulative distribution function (bottom panels) of merger statistic $S$ at $z=0.5$ (left), $z=1$ (middle), and $z=2$ (right) for all galaxies (black), galaxies which have hosted high-mass (red) or low-mass (blue) major BH merger in the past 300 Myr.  The error bars show the mean and standard deviation of each sample.  For most BH mergers, the distribution of the $S$-statistic is relatively unchanged compared to the general distribution for all galaxies.  For high-mass major BH mergers occurring within the past $\sim 500$~Myr (and especially within 300 Myr) the galaxies tend to show morphological evidence of having recently merged.  The dashed red and blue curves show the weighted distribution of the full galaxy population to provide a mass-matched sample to the merger population (i.e. match of the $M_{\rm{tot}}$ distribution of the red and blue populations, respectively). }
    \label{fig:1d_histograms}
\end{figure*}

\begin{figure}
    \centering
    \includegraphics[width=0.49\textwidth]{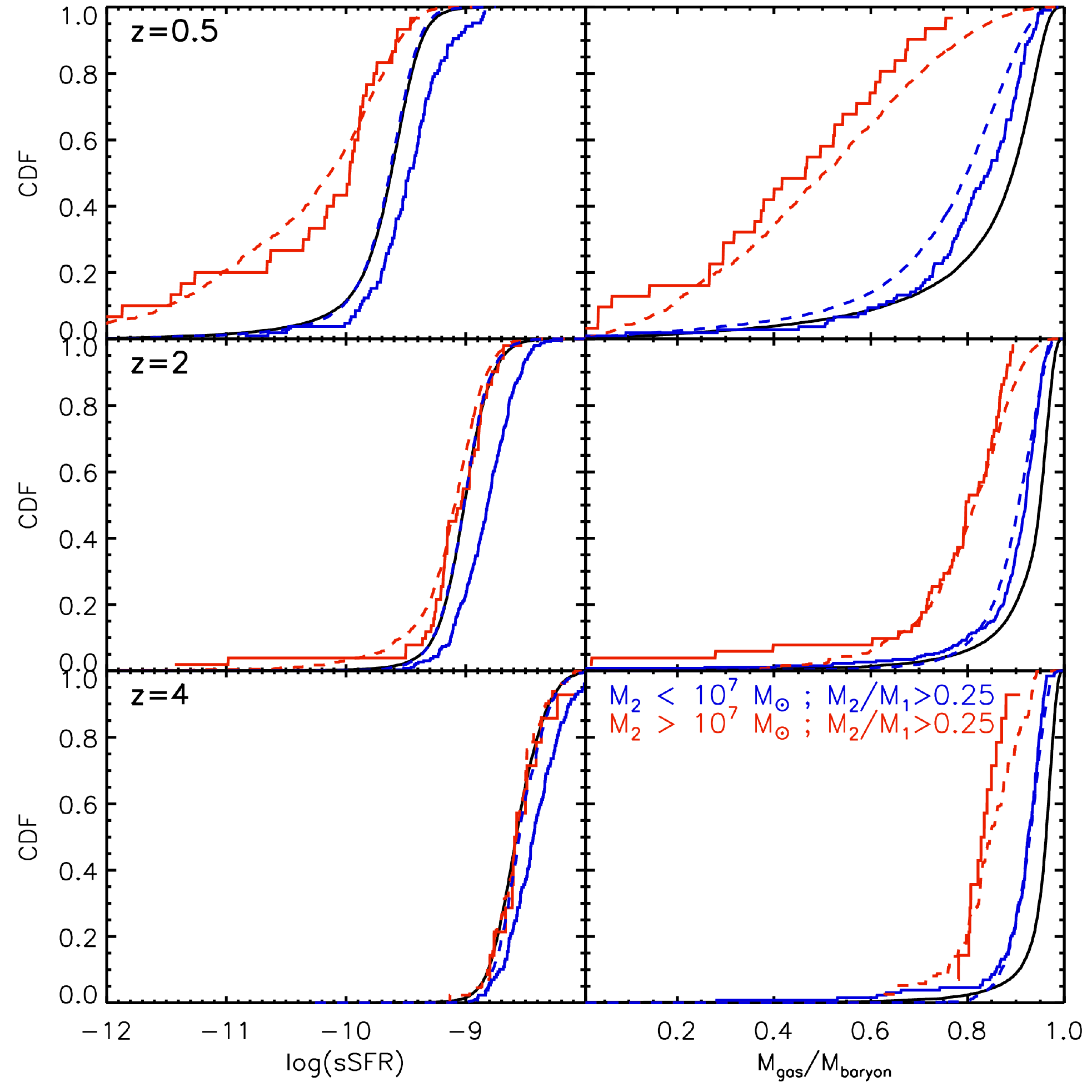}
    \caption{Cumulative distribution function of sSFR (left) and $M_{\rm{gas}}/M_{\rm{baryon}}$ (right).  Black lines show the distribution for the full galaxy population.  Solid coloured lines show a sample of galaxies hosting BH mergers.  Dashed coloured lines show the distribution of the full galaxy population mass-weighted to match the distribution of the targeted BH merger sample.  Hosts of low-mass major BH mergers tend to have slightly increased sSFR (i.e. SF boost from the galaxy merger).  Hosts of high-mass major BH mergers do not exhibit such a boost.  At low-$z$, the low gas fraction could explain the lack of SFR boost; at high-$z$, however, the galaxies are gas-rich, suggesting another factor must play a role, such as feedback from the high-mass BHs. }
    \label{fig:ssfr_gasfraction}
\end{figure}

We confirm this in Fig.~\ref{fig:ssfr_gasfraction}, where we show the cumulative distribution function for sSFR (left column) of the full population (black) compared to the major BH merger populations (red and blue).  Given the strong correlation between sSFR and galaxy mass (Fig.~\ref{fig:ssfr_mtot}), it is necessary to compare mass-matched samples rather than comparing with the full galaxy population.  As in Fig.~\ref{fig:1d_histograms}, for each target sample (line colour) we show the distribution for a mass-matched equivalent from the full galaxy sample (dashed lines).  Here we find that the hosts of high-mass major BH mergers (solid red) are equivalent to the mass-matched sample from the full population (dashed red).  This suggests that hosting a major massive BH merger either tends to have minimal impact on the sSFR, or the effect is short compared to the $300$~Myr used for selecting the host sample.  Conversely, we find that hosts of low-mass major BH mergers tend to have slightly higher sSFR than the mass-matched sample, suggesting that the galaxy merger triggers an increase in SFR (though only on the order of $\sim 0.5\sigma$).  We find qualitatively similar results across redshifts: higher redshifts have higher sSFRs overall, but high-mass major BH merger hosts match an equivalent-mass sample whereas low-mass major BH merger hosts tend to have slightly higher sSFR than their equivalent-mass sample.

One possible explanation for why low-mass galaxy mergers may boost the sSFR, while high-mass galaxy mergers do not, is simply due to the gas richness of the merging galaxies. High-mass galaxies, where the high-mass major BH mergers tend to occur, generally have low gas-fractions, while the lower-mass galaxies which host lower-mass BH mergers would be expected to be more gas-rich, and thus have more potential to drive a burst of star formation \citep[e.g.][]{Vogelsberger2014b, Genel2014}.  Indeed, we find this to be the case at $z=0.5$: galaxies hosting high-mass major BH mergers tend to have $M_{\rm{gas}}/M_{\rm{b}} \sim 0.43 \pm 0.22$, compared to low-mass major BH merger hosts, which have $M_{\rm{gas}}/M_{\rm{b}} \sim 0.80 \pm 0.16$ (see right panels of Fig.~\ref{fig:ssfr_gasfraction}).

However, we note that this cannot be the sole explanation. In the middle and lower panels of Fig.~\ref{fig:ssfr_gasfraction} we show the sSFR and gas fractions at $z=2$ and $z=4$.  Similar to low-$z$, the high-mass major BH merger hosts have the same sSFR as a mass-matched sample, while the low-mass major BH merger hosts tend to have slightly higher sSFR than their mass-matched sample. However, the gas fractions are much higher, with the high-mass hosts at $z=2$ having higher gas fractions than the low-mass hosts at $z=0.5$.  Thus, at high-$z$ we have gas rich galaxy mergers which still do not appear to affect the star formation of galaxies hosting high-mass major BH mergers, while there is a weak correlation for low-mass major BH mergers (sSFR is typically $\sim 0.5 \sigma$ higher among the low-mass major BH merger hosts). This suggests that the black holes themselves may play a role in preventing the star formation from increasing.  In particular, we know that feedback from massive black holes can self-regulate \citep[e.g.][]{King2003, Springel2005, Costa2014, Sijacki2015}, and can suppress SF, 
while the low mass black hole may not be capable of providing sufficient feedback to do so.  Indeed, we find a correlation between the energy emitted by the black hole and the position of the galaxy on the star forming main sequence several hundred Myr later, with higher AGN feedback during the merger corresponding to lower sSFR post-merger. This provides additional evidence that feedback from sufficiently massive black holes can help regulate star formation following a major galaxy merger, suppressing the boost in SFR one might otherwise expect from a gas-rich galaxy, as shown in Fig.~\ref{fig:evo1}.

\section{Merger Timescales}
\label{sec:mergerdelays}

Having found evidence that galaxies show morphological signatures of a recent galaxy merger over timescales on the order of 300-500 Myr, we now consider the typical BH merger timescales and the impact that may have on our results.  Within the Illustris simulation (and indeed many similar simulations), a pair of black holes merge as soon as their separation is less than the particle's smoothing length, rather than incorporating a coalescence time for the binary.  Recently, several works have attempted to estimate the expected coalescence time for binary black holes by post-processing cosmological simulations, finding timescales on the order of 100s of Myrs to Gyrs for the binary coalescence time \citep[e.g.][]{Blecha2016, Kelley2017b, Rantala2017, Mannerkoski2019, Sayeb2020}.  Additionally, the time for a satellite black hole to reach the center of a galaxy and form a binary with the central black hole can also be on the order of 100s of Myrs to Gyrs, based on the dynamical friction timescale for infall to the galactic center \citep[e.g.][]{Volonteri2020}. The galaxy structure, e.g. the existence or lack of a dense stellar core, can affect the time satellite black holes spend at large radii \citep{Tremmel2018a, Tremmel2018b, Barausse2020}, and directly incorporating dynamical friction into cosmological volumes  suggests that the orbital decay timescale may be both substantial and redshift-dependent \citep{Bartlett2021}.  The Illustris simulation uses a re-centering scheme whereby black holes are re-positioned toward the local potential minimum; this prevents numerical wandering of black holes, but also means the the full infall time to reach the galaxy center may be notably under-estimated.
Furthermore, both simulations \citep[e.g.][]{Bellovary2019} and observations \citep[e.g.][]{Reines2020} suggest that black holes in dwarf galaxies may frequently be located offset from the galaxy center, which could further delay any mergers involving low-mass black holes seeded into Illustris (which are initially placed at the galaxy center).
Overall, this suggests that the Illustris simulation likely overestimates the speed with which BHs merge following the merging of their host galaxies, and properly accounting for this has the potential to impact the expected GW detection rate, shift the peak detection time to lower redshift, and prevent GW hosts from being visibly disturbed. Although a complete investigation into accurately estimating the time delays remains beyond the scope of this paper, we address this, by imposing a delay between when the BH particles merge in the simulation (which is closer to when the BH binary may form) and when the final coalescence and GW emission occur.

\begin{figure}
    \centering
    \includegraphics[width=0.49\textwidth]{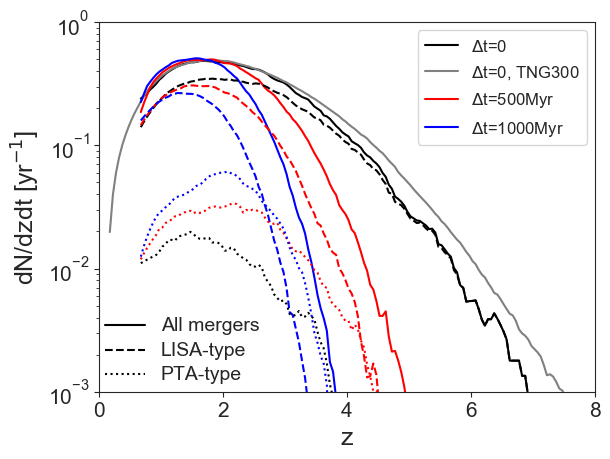}
    \caption{\textit{Solid lines:} The expected BH merger rate signal as a function of redshift.  In black we show the rates from the original simulation (with a comparison to TNG300 in grey), while in red and blue we show the rates if we were to assume a fixed delay time of 500 or 1000 Myr, respectively.  Incorporating a delay time has a significant impact on the earliest BH mergers, but the low-redshift merger rate is largely unaffected by a delay time. \textit{Dashed/dotted lines:} The BH merger rate signal for mergers corresponding to PTA-type events (dotted; defined as $M_{\rm{chirp}} > 10^8\, {\mathrm M_\odot}$) and LISA-type events (dashed; defined as $M_{\rm{chirp}} < 10^7\, {\mathrm M_\odot}$). Note that the rate of LISA-type mergers is limited to those resolved by the Illustris simulation, and that the rate of LISA-detectable mergers is expected to be higher (see main text for further details).  Thus incorporating a BH merger delay can significantly change the merger rates, with a merger delay not only decreasing the rate of high-$z$ mergers (especially at low-mass), but also significantly increasing the rate of high-mass mergers at $z \sim 2$ (near the merger rate peak).}
    \label{fig:dndt_z}
\end{figure}

In Fig.~\ref{fig:dndt_z}, we plot the predicted rate of supermassive black hole mergers throughout the observed universe as a function of merger redshift (solid black line), in terms of the rate at which the signals would reach Earth:
\begin{equation}
    \frac{{\rm d}N}{{\rm d}z\,{\rm d}t} = \frac{1}{z_2-z_1} \int_{z_1}^{z_2} \frac{{\rm d}^2 n(z)}{{\rm d}z\,{\rm d}V_c} \frac{{\rm d}z}{{\rm d}t} \frac{{\rm d}V_c}{{\rm d}z} \frac{{\rm d}z}{1+z}\,.
\end{equation}
The first BH merger in the simulation occurs at $z \sim 7.7$ (though we note this is limited by the boxsize, resolution and seed criteria used), and the rate then grows with redshift to a peak at $z \sim 2$, followed by a decline at later times, broadly consistent with similar analyses \citep[e.g.][]{Salcido2016, Katz2020}.  We also note that the BH merger rate in Illustris (black) is consistent with that from the TNG300 simulation (grey), except that the larger-volume TNG300 has slightly more BH mergers at very high redshift.  In addition to the Illustris predictions, we also use the larger-volume Illustris TNG300 simulation \citep{Springel2018} to get predicted BH merger rates (grey line).  The larger boxsize provides a less noisy redshift distribution, and very slightly higher rates at the highest redshift.  Otherwise, the two simulations are largely equivalent, as we would expect given the similar seeding and merging conditions between the two simulations.  

In addition to the overall rate, we also show the BH merger rates if the expected inspiral times for the mergers were 500 Myr (red lines) or 1 Gyr (blue lines) following when in Illustris a given black hole pair merged (which did not incorporate any expected inspiral/hardening time).  Here we use fixed time delays for all BH mergers as a means to show the typical impact which a true merger timescale may have; a more accurate investigation using variable time delays based on the BH and galaxy properties is beyond the scope of this work, but will be tested in future work.  The incorporation of a delay time necessarily has a strong impact on the very early BH mergers.  However, we find that at lower redshifts the total BH merger rate evolves slowly enough that adding a delay on the order of 1 Gyr has essentially no impact on the total rate at which we can expect mergers to take place.  This is consistent with \citet{Weinberger2018}, who showed that in the Illustris TNG300 simulation, BH merger rates binned by mass and redshift have a relatively minor evolution with redshift, and thus concluded that implementing a more realistic inspiral time would not significantly affect the global merger rate.

\begin{figure}
    \centering
    \includegraphics[width=0.49\textwidth]{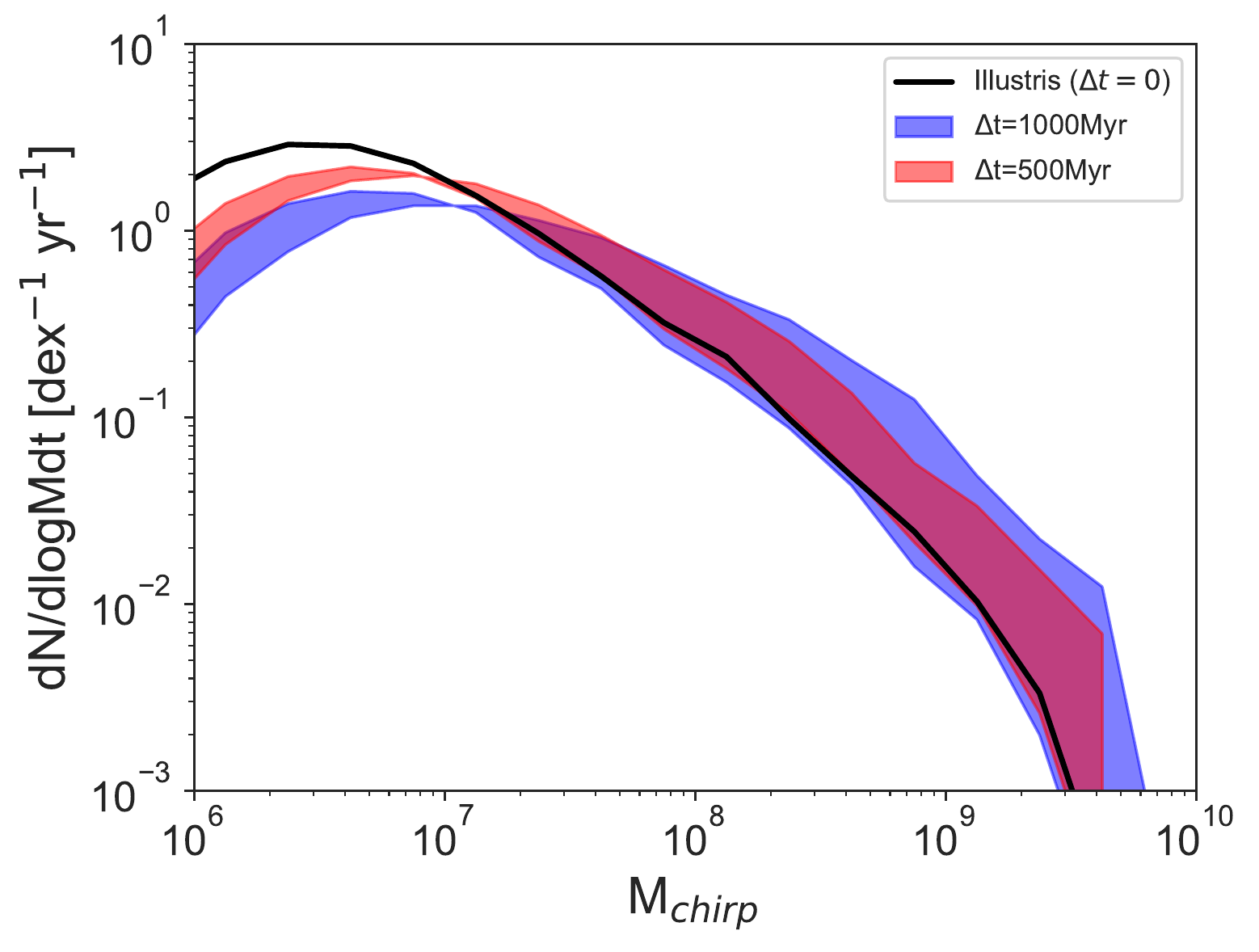}
    \caption{The predicted BH merger rate as a function of chirp mass for the original simulation and after adding 500 and 1000 Myr delay times.  For a given $\Delta t$, the shaded region spans the range of possible values depending on the accretion efficiency of the binary black holes (between Bondi- and Eddington-like growth; see text for details). }
    \label{fig:dndt_chirp}
\end{figure}

However, the inspiral time is more than merely a delay on the time at which the BH merger occurs; it also gives the progenitor black holes more time to grow prior to the merger, so we would expect the merger masses to increase with longer delay times (especially as galaxy mergers can often be followed by periods of efficient growth as gas rich galaxy mergers drive additional gas into the central region where the black holes are accreting). We show the expected effect this will have with dashed and dotted lines in Fig.~\ref{fig:dndt_z}, where we consider the rate of merging black holes with $M_{\rm{chirp}} < 10^7\, {\mathrm M_\odot}$ (dashed, roughly corresponding to LISA-detectable mergers) and $M_{\rm{chirp}} > 10^8\, {\mathrm M_\odot}$ (dotted; roughly corresponding to PTA-detectable mergers).  We note here that the rate of LISA-type mergers is limited to those BH mergers resolved by the Illustris simulation, which is limited by the seed mass of $M_{\rm{seed}} = 10^5 \: h^{-1} \: M_\odot$, whereas the actual detection range of LISA should peak at lower-mass scales \citep[see, e.g.][]{Ricarte2018, Dayal2018, DeGraf2019}; thus the LISA-type merger curve shows how incorporating a delay can impact the rate of the higher-mass LISA detections, but should nonetheless represent a lower-limit for the full detection rate.

In the case of the $\Delta t > 0$ curves, we assume that the growth of the inspiraling black holes during that delay time is at the equivalent Eddington fraction as the post-merged black hole in the original simulation (i.e. we check the fractional accretion in Illustris of the merged black hole over the course of $\Delta t$, and let each progenitor black hole grow by the same fraction).  Here we see that in addition to decreasing the number of high-z BH mergers (especially for low-mass, LISA-type mergers), longer delay times tend to result in more high-mass BH mergers at lower-$z$ as the inspiralling black holes have more time to grow prior to merging.  In particular, we find that longer merger timescales which include the associated black hole growth lead to an increase in the rate of high-mass (PTA-type) mergers at $z \lesssim 3.5$, especially near the peak at $z \sim 2$.

\begin{figure*}
    \centering
    \includegraphics[width=0.99\textwidth]{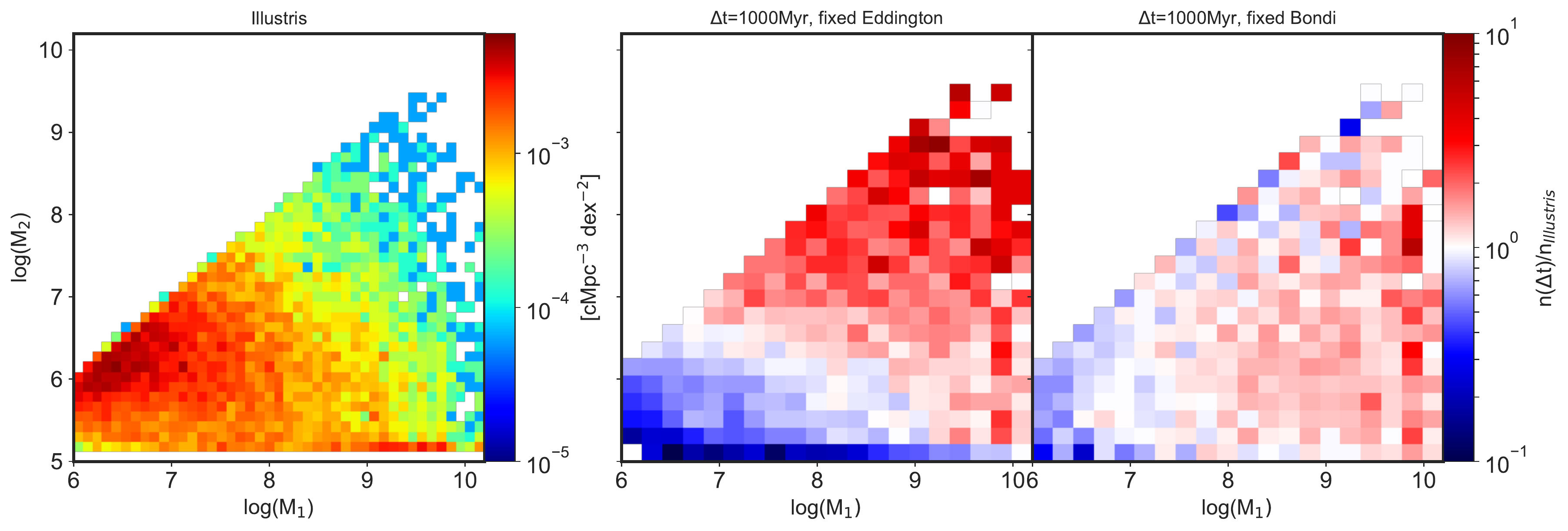}
    \caption{\textit{Left panel:} The distribution of primary ($M_1$) and secondary ($M_2$) BH merger masses in the Illustris simulation, coloured by the number density of mergers within the simulation.   \textit{Middle and right panels:} The ratio between BH mergers in the original simulation ($n_{\rm{Illustris}}$) and after imposing a 1 Gyr delay on all mergers ($n(\Delta t)$), assuming growth during the specified $\Delta t$ continues at the equivalent Eddington rate (middle panel) or equivalent Bondi rate (right panel; see text for details).  If growth is Eddington-dependent, both pre-merged black holes grow equivalently, decreasing the number of low-mass mergers with a corresponding increase in high-mass mergers (as each merger moves along a line with slope of 1 in the $M_2 - M_1$ plane).  If growth is Bondi-dependent, then the number of low-mass mergers decreases, but most of the growth occurs in the more massive black hole, such that the main shift is in the $M_1$-direction (i.e. slope less than one).  The colour scale shows the number density of BH mergers after imposing a 1 Gyr delay relative to the number density of mergers in the original simulation; red represents an increase in the number of mergers at the given mass scales, while blue shows a decrease.}
    \label{fig:mergermasses}
\end{figure*}

\begin{figure}
    \centering
    \includegraphics[width=0.49\textwidth]{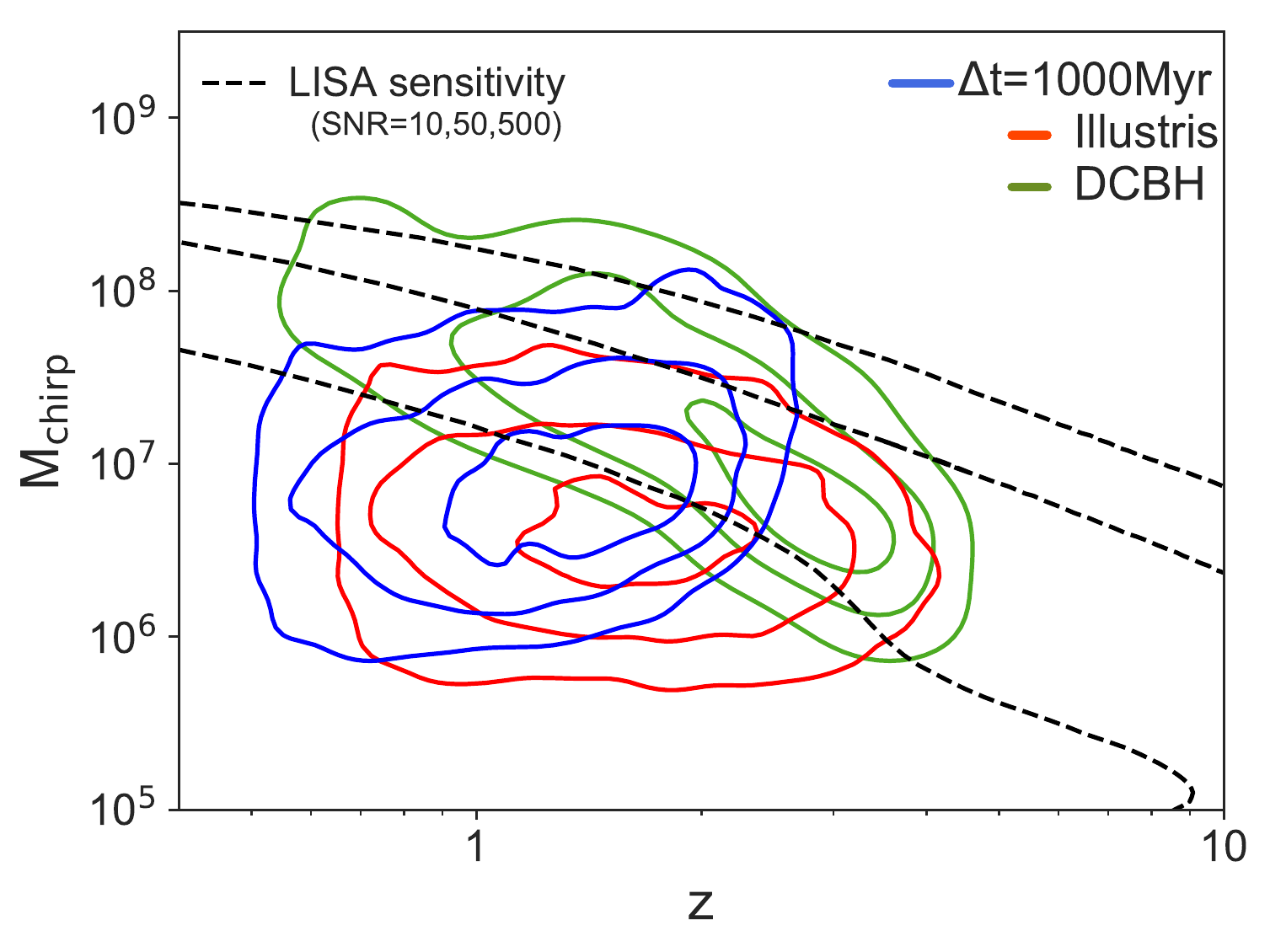}
    \caption{The redshift - chirp mass distribution of BH mergers in the original Illustris simulation (red contours) and after applying a 1 Gyr delay (blue contours), with contour levels encompassing $\sim 82 \%, \sim 55 \%$, and $\sim 15 \%$ of mergers in each case.  Typical chirp masses are largely unaffected by delay times at low redshift, but significantly increased in high-$z$ mergers.  This is in contrast to the expected impact from changing the seed formation model to only consider direct collapse black holes, which predicts chirp mass to increase with time \citep[green contours; see][]{DeGraf2019}. }
    \label{fig:mchirp_vs_z}
\end{figure}

We further consider the effect this may have in Fig.~\ref{fig:dndt_chirp}, where we show the BH merger rate as a function of chirp mass ($M_{\rm{chirp}}=(M_1 M_2)^{3/5} / (M_1 + M_2)^{1/5}$), both from the original simulation (black) and for inspiral time delays of $\Delta t = 500$~Myr (red) and $\Delta t = 1$~Gyr (blue).  For these time delay cases, we consider two growth efficiencies, where the inspiral growth matches either the Eddington fraction or the Bondi fraction of the post-merged black hole from the original simulation.  

We expect that the assumption of a fixed Eddington-fraction to represent a likely upper bound, as the higher mass from the original simulation (i.e. the combined post-coalescence mass) will tend toward a higher efficiency than the lower mass from the pre-merged black hole pair.  Conversely, although the simulation uses a Bondi formalism to model accretion, the fixed Bondi rate likely represents an underestimate for the gas accretion, as this assumes each black hole is isolated, rather than the combined binary we have here.  As such, using both fixed-Eddington and fixed-Bondi fractions provides a range of accretion efficiencies which span a good fraction of the uncertainty within our assumptions.

Here we again see that although the overall BH merger rate is largely unchanged, the merger masses are affected by the inspiral time, with the longer delays resulting in more massive BH mergers: the mergers peak at $M_{\rm{chirp}} \sim 2 \times 10^6\, {\mathrm M_\odot}$ in the original simulation, $\sim 7 \times 10^6\, {\mathrm M_\odot}$ for a 500~Myr delay, and $\sim 10^7\, {\mathrm M_\odot}$ for a 1~Gyr delay. Furthermore, the delay has the potential to increase the number of massive BH mergers by up to an order of magnitude, depending on the efficiency of growth during the binary phase.  However, we again note that the low-mass end is limited by the seed mass of black holes in the Illustris simulation, and would thus expect the actual detection rate for LISA to be significantly higher \citep[see, e.g.][]{Ricarte2018, Dayal2018, DeGraf2019}.

We show the distribution of BH merger masses in more detail in Fig.~\ref{fig:mergermasses}: in the left panel we plot the distribution of the primary ($M_1$) and secondary ($M_2$) merger masses in Illustris, finding that mergers peak at $M_1 \sim 5 \times 10^6\, {\mathrm M_\odot}$ and $M_2 \sim 10^6\, {\mathrm M_\odot}$.  We also show the fractional change in merger masses when including a delay ($n(\Delta t=1\,{\rm Gyr}) / n(\Delta t = 0)$), assuming growth at a fixed Eddington fraction (middle panel) or fixed Bondi fraction (right panel).
In both growth models, we find that incorporating a delay time decreases the number of low-mass mergers while increasing high-mass mergers, but with several key differences.  The more efficient growth in the fixed Eddington model results in a larger change in the BH merger rate, by as much as 1 dex more high-mass mergers and 1 dex fewer low-mass mergers.  However, we note that by construction the fixed Eddington growth model does not affect the mass ratio of the BH mergers, since fixed Eddington fraction growth means both black holes grow proportionally during the binary phase.  In contrast, assuming growth based on a fixed Bondi fraction has a smaller overall impact, but does affect the mass ratios.  In particular, Bondi growth rates scale with $M_{\rm{BH}}^2$, so incorporating a delay assuming Bondi efficiency means that high mass-ratio BH mergers actually feature increasing mass ratios with longer delays (since the less massive progenitor black hole is less able to grow during the delay time), while the mass ratio of major mergers are relatively unaffected (since both progenitors grow equally efficiently).  

The net result of Figs.~\ref{fig:dndt_z}-\ref{fig:mergermasses} is that a non-zero BH merger timescale has the potential to significantly impact the detection rate (both at LISA and PTA sensitivities), and the chirp mass and mass ratios of detectable mergers.  However, the magnitude of the changes remains sensitive to the precise accretion prescription during the binary phase, which cannot be fully addressed in a post-processing analysis.  Rather, this demonstrates the need for high-resolution simulations which directly incorporate a hardening timescale and the associated accretion onto each black hole of the binary.

In Fig.~\ref{fig:mchirp_vs_z}, we plot the typical chirp masses as a function of redshift for the original simulation (red contours) and after incorporating a delay (blue contours).  As in Fig.~\ref{fig:dndt_z}, we find the largest effect at the earliest times. High redshift merging BH masses are most strongly impacted by the inspiral time: combining the efficient growth of high-$z$ black holes with a long inspiral time relative to the current black hole age leads to a dramatic increase in typical chirp masses.  In contrast, at low redshift we see only a relatively minor increase in the typical masses.  This impact on typical chirp masses lies within LISA sensitivity (dashed lines), and so is necessarily when considering expected LISA detections.  However, we note that the low-mass end of the contours here are again limited by the simulation resolution, and that LISA should detect significant numbers of BH mergers below the minimum chirp mas probed by Illustris.
We also show the evolution of chirp mass for black holes seeded assuming direct collapse black hole (DCBH) seed formation, using the results of \citet{DeGraf2019}, which show an increasing $M_{\rm{chirp}}$ with time.  This time-evolution is due to the change in seed criteria: the direct collapse conditions (especially the metallicity requirement) are generally more easily satisfied at high redshift. Thus, direct collapse seeds in our simulations tend to form at high redshift and grow relatively efficiently, resulting in few low-mass black holes at low redshift (though other formation pathways may produce additional low-mass low redshift BHs).  Both these additions to the BH model (incorporating a coalescence time, and modeling seed formation through a more physically motivated direct collapse model) result in increasing the expected merging BH masses.  However, as we see in Fig.~\ref{fig:mchirp_vs_z}, the redshifts at which the typical masses increase is reversed: modeling formation via direct collapse leads to an increase of  $M_{\rm{chirp}}$ with time, whereas incorporating a coalescence time tends to result in a decreasing $M_{\rm{chirp}}$ with time.  Thus, we note that measuring the redshift evolution of the chirp mass measurement will provide an important means of differentiating between impact of seed formation vs. BH merger timescales.

\section{Conclusions}
\label{sec:conclusions}

In this work, we have taken advantage of the cosmological galaxy formation simulation Illustris to investigate the mergers of SMBHs and the connection between BH mergers and the morphologies of the galaxies in which they are found. Our main results are as follows.

\begin{itemize}

\item Gravitational wave detections of supermassive black hole mergers should probe a large range of BH masses, with PTAs probing the massive end of the $M_{\rm{BH}}-M_*$ relation and LISA's sensitivity probing the low mass end of the $M_{\rm{BH}}-M_*$ relation (beyond that resolved by the Illustris simulation).  Most BH mergers involve the central black hole of a given galaxy, with $M_{\rm{chirp}}$ being within $\sim$0.5 dex of $M_{\rm{BH,cen}}$. 

\item By tracking several illustrative BH mergers in time, we find that for $z \lesssim 1$, galaxies hosting massive major BH mergers frequently show morphological signatures of a recent galaxy merger.  Prior to BH mergers, galaxies are generally undisturbed.  After the merger, however, we find galaxies showing multiple distinct cores (i.e. in-process galaxy merger), disturbed stellar components, tidal tails, and shells of stellar light, indicative of a recent galaxy merger, though some galaxies maintain morphologies lacking direct evidence for a recent galaxy merger.

\item We then statistically characterize galaxies as having morphological evidence of merging using the Gini-$M_{20}$ relation.  Minor BH mergers hosts do not tend to show galaxy merger morphologies, and low-mass major BH mergers only exhibit short-lived merger morphologies at $z \sim 0.5$.  However, high-mass major BH mergers tend to be found in galaxies which display merger morphologies.  The correlation between BH mergers and host morphology is strongest at $z \sim 0.5$, and the morphological signal tends to survive for of the order of $\sim$500~Myr.  

\item Although the majority of BH merger hosts show disturbed morphologies at $z \sim 0.5$, BH merger hosts represent a minority of disturbed galaxies. Hence, we would expect a BH merger to occur in a galaxy with a disturbed host, but we would not expect most disturbed galaxies to host a recent BH merger with a detectable gravitational wave signal.

\item Other than morphology, BH merger hosts tend to have similar properties to non-merger hosts, with the exception of star formation.  Galaxies hosting low-mass BH mergers tend to have a slightly higher sSFR compared to a mass-matched sample, suggesting that the galaxy merger triggers a burst of star formation.  In contrast, galaxies hosting high-mass BH mergers have equivalent sSFR to a mass-matched sample.  At low-$z$, these high-mass galaxies tend to have low gas fractions, and thus we would not expect to see an increase in SFR as a result of a dry galaxy merger.  At high-$z$, the gas fractions are quite high, suggesting that the galaxy merger should trigger an increase in SF.  However, this would also trigger increased BH accretion and associated AGN feedback, which can suppress SF, counteracting the boost that would otherwise be found.

\item The $\sim$500 Myr survival time for morphological evidence in a galaxy following a high-mass major BH merger should include the inspiral/hardening time of the BH binary which is however not modelled in our simulations. Incorporating a more realistic time for a satellite black hole to fall to the galactic center and then for the BH binary to merge would suggest that the host galaxy will have time to relax prior to the emission of gravitational waves.  As such, it may not be possible for electromagnetic followups to a GW detection to be able to morphologically identify the host galaxy of the BH merger.  However, this is sensitive to the time it takes before the GW signal is emitted, and thus confirmation will require new simulations which directly incorporate physically realistic BH merger delay times.

\item A simple post-processing addition of a delay time to all BH mergers has minimal impact on the global BH merger rate at low redshift. However, the delay can affect the masses of the merging black holes, resulting in higher-mass BH mergers overall.  In particular, the number of low mass BH mergers decreases, while the number of high-mass major BH mergers increases, which results in a typical BH merger chirp mass which increases at high-$z$ (compared to the $z$-independent chirp mass in the original Illustris simulation).

\item Accurately incorporating the BH merger timescale may impact the rate at which we expect to detect gravitational waves. In particular, our analysis suggests that a longer delay time may decrease the number of LISA-detections, while increasing the number of detections from PTAs.  However, this is highly sensitive to the accretion efficiency of the black holes during the infall and binary phase, further emphasizing the importance of performing simulations which self-consistently incorporate BH merger times and the BH growth which occurs during those times.

\end{itemize}

In summary, we have found that supermassive black hole mergers occur in galaxies with typical $M_{\rm{BH}}-M_*$ ratios, though the morphologies are often disturbed.  In particular, we find that the galaxies hosting high-mass major BH mergers show morphological evidence of a recent galaxy merger, which generally survives for $\sim 500$ Myr.  However, this survival timescale should include the inspiral/hardening times which the Illustris simulation does not incorporate. The fact that the BH merger delay time should be on the same order suggests that by the time a gravitational wave from a BH merger is emitted, the morphological evidence in the host galaxy may have been lost.  However, this is sensitive to the BH merger delay timescale which is highly dependent on the local medium properties around the binary, emphasizing the need to run further simulations which directly incorporate a realistic model for merger timescales, to characterize the fraction of galaxies whose morphology may survive until black hole coalescence.  

In addition to the correlation to host morphology, we also showed that although a reasonable delay for merging black holes does not dramatically affect the total number of BH mergers, it has the potential to significantly impact the masses when they merge.  This is particularly important for upcoming GW surveys, as the additional growth during the binary black hole phase may decrease the rate of detections from LISA, increase the detections from PTAs, and shift the distribution of merging BH masses from both toward higher masses which increase with redshift.  The magnitude of these shifts depends on the accretion efficiency of both the binary black hole system as a whole and the relative efficiency of each component black hole. Hence, next generation cosmological simulations able to realistically track supermassive black hole binary inspirals as well as accretion flows onto the binary will play a key role in making quantitatively accurate predictions for upcoming gravitational wave surveys.

\section*{Acknowledgments}
We would like to thank Marta Volonteri and Alberto Sesana for useful and interesting discussions which benefited this work, and the anonymous referee whose comments improved this paper.  DS would like to thank Prof. D’Anchise for all the help and support in finalizing this manuscript. CD and DS acknowledge the support from the ERC starting  grant 638707 ``Black holes and their host galaxies: co-evolution  across  cosmic time'' and the STFC. This research used: The Cambridge Service for Data Driven  Discovery  (CSD3),  part  of  which is operated by the University  of  Cambridge  Research  Computing  on  behalf of the STFC DiRAC HPC Facility (www.dirac.ac.uk).  The DiRAC component of CSD3 was funded by BEIS capital  funding  via  STFC  capital  grants  ST/P002307/1  and ST/R002452/1 and STFC operations grant ST/R00689X/1.  DiRAC  is  part  of  the  National  e-Infrastructure.  Simulations were run on the Harvard Odyssey and CfA/ITC clusters, the Ranger and Stampede supercomputers at the Texas Advanced Computing Center as part of XSEDE, the Kraken supercomputer at Oak Ridge National Laboratory as part of  XSEDE,  the  CURIE  supercomputer  at  CEA/France  as part of PRACE project RA0844, and the SuperMUC computer at the Leibniz Computing Centre, Germany, as part of project pr85je.  TDM acknowledges funding from NSF ACI-1614853,  NSF AST-1616168, NASA ATP 19-ATP19-0084, NASA ATP 80NSSC18K101, and NASA ATP NNX17AK56G.
We also acknowledge NASA ATP 80NSSC20K0519.

\section*{Data Availability}
The data underlying this article were derived from sources in the public domain: The Illustris Project (https://www.illustris-project.org/) and The IllustrisTNG Project (https://www.tng-project.org/)

\bibliographystyle{mn2e}       
\bibliography{astrobibl}       

\end{document}